\documentclass[sigconf,screen]{acmart}

\usepackage{algorithmic}
\usepackage{graphicx} 
\usepackage{textcomp} 
\usepackage{xcolor}
\usepackage{listings}
\usepackage[capitalise]{cleveref}
\usepackage{comment}
\usepackage{multirow}
\usepackage{enumitem}
\usepackage{subcaption}
\usepackage{stfloats}
\usepackage{float}
\usepackage{soul}
\usepackage[labelfont=bf,textfont=bf]{caption}
\DeclareCaptionFont{ninept}{\fontsize{9pt}{11pt}\selectfont #1}
\usepackage{pifont}
\usepackage{array}
\usepackage{booktabs}
\usepackage{fontawesome}
\usepackage{color}
\usepackage{xcolor}
\usepackage{colortbl}
\usepackage{makecell}

\usepackage{fancyhdr} 

\AtBeginDocument{%
  }

\copyrightyear{2025}
\acmYear{2025}
\setcopyright{cc}
\setcctype{by}
\acmConference[ISCA '25]{Proceedings of the 52nd Annual International Symposium on Computer Architecture}{June 21--25, 2025}{Tokyo, Japan}
\acmBooktitle{Proceedings of the 52nd Annual International Symposium on Computer Architecture (ISCA '25), June 21--25, 2025, Tokyo, Japan}\acmDOI{10.1145/3695053.3731029}
\acmISBN{979-8-4007-1261-6/2025/06}

\newcolumntype{b}{X}
\newcolumntype{s}{>{\hsize=.7\hsize}X}
\newcolumntype{z}{>{\hsize=1.3\hsize}X}
\let\oldding\ding
\renewcommand{\ding}[2][1]{\scalebox{#1}{\oldding{#2}}}

\usepackage{comment}

\usepackage{lipsum}

\newcommand{\acol}{orange}
\newcommand{\bcol}{cyan}
\newcommand{\ccol}{blue}
\newcommand{\dcol}{purple}
\newcommand{\ecol}{teal}

\usepackage{xparse}
\usepackage{ifthen}

\newif\ifshowreviewer
\showreviewerfalse

\NewDocumentCommand{\mf}{m}{%
  \ifshowreviewer
    \begingroup
    \setlength{\fboxsep}{1pt}%
    \IfEqCase{#1}{%
      {reva}{{\color{\acol} {\bf \small \fbox{RevA}}}}%
      {revb}{{\color{\bcol} {\bf \small \fbox{RevB}}}}%
      {revc}{{\color{\ccol} {\bf \small \fbox{RevC}}}}%
      {revd}{{\color{\dcol} {\bf \small \fbox{RevD}}}}%
      {reve}{{\color{\ecol} {\bf \small \fbox{RevE}}}}%
    }[{{\color{olive} \bf \small \fbox{#1}}}]%
    \endgroup
  \else
  \fi
}

\newcommand\typo[1]{\noindent{{#1}}} 
\usepackage[normalem]{ulem} 

\makeatletter
\newcommand{\rv}[2]{%
  \mf{#1}
  \hypertarget{#2}{\label{#2}}
}
\newcommand{\rvc}[2]{%
  \hypertarget{#2}{\label{#2}}
}

\makeatother

\newcommand{\ttt}[1]{\begingroup\spaceskip=0.5\fontdimen2\font plus 0.5\fontdimen3\font minus 0.5\fontdimen4\font\noindent\texttt{#1}\endgroup}
\newcommand\workname[1]{\begingroup\spaceskip=0.5\fontdimen2\font plus 0.5\fontdimen3\font minus 0.5\fontdimen4\font\noindent\textit{#1}\endgroup}

\definecolor{codegray}{rgb}{0.5,0.5,0.5}
\definecolor{codegreen}{rgb}{0,0.6,0}
\definecolor{codeblue}{rgb}{0,0,0.9}
\definecolor{backcolour}{rgb}{0.95,0.95,0.95}

\lstdefinestyle{mystyle}{
    commentstyle=\color{codegreen},
    keywordstyle=\color{codeblue},
    stringstyle=\color{codeblue},
    basicstyle=\normalsize\ttfamily,
    breakatwhitespace=false,         
    breaklines=true,                 
    captionpos=b,                    
    keepspaces=true,                 
    showspaces=false,                
    showstringspaces=false,
    showtabs=false,                  
    tabsize=2
}
\begin{document}


\title[Garibaldi: Instruction-Data Pairwise Shared Last-Level Cache Management in Server Workloads]{Garibaldi: A Pairwise Instruction-Data Management for Enhancing Shared Last-Level Cache Performance in Server Workloads}



\author{Jaewon Kwon}
\authornote{These two authors contributed equally to this work.}

\orcid{0009-0000-6352-0546}
\email{jaewon.kwon@yonsei.ac.kr}
\affiliation{%
  \institution{Yonsei University}
  \city{Seoul}
  \country{Republic of Korea}}

\author{Yongju Lee}
\authornotemark[1]
\orcid{0009-0008-3426-9426}
\email{yongju.lee@yonsei.ac.kr}
\affiliation{%
  \institution{Yonsei University}
  \city{Seoul}
  \country{Republic of Korea}}

\author{Jiwan Kim}
\orcid{0009-0008-4738-9240}
\email{jiwan.kim@yonsei.ac.kr}
\affiliation{%
  \institution{Yonsei University}
  \city{Seoul}
  \country{Republic of Korea}}

\author{Enhyeok Jang}
\orcid{0009-0000-7034-6793}
\email{enhyeok.jang@yonsei.ac.kr} 
\affiliation{%
  \institution{Yonsei University}
  \city{Seoul}
  \country{Republic of Korea}}

\author{Hongju Kal}
\orcid{0009-0001-8443-1734}
\email{hongju.kal@yonsei.ac.kr}
\affiliation{%
  \institution{Yonsei University}
  \city{Seoul}
  \country{Republic of Korea}}

\author{Won Woo Ro}
\orcid{0000-0001-5390-6445}
\email{wro@yonsei.ac.kr}
\affiliation{%
  \institution{Yonsei University}
  \city{Seoul}
  \country{Republic of Korea}}














\received{22 November 2024}
\received[revised]{21 February 2025}
\received[accepted]{22 March 2025}




\begin{abstract}


Modern CPUs suffer from the frontend bottleneck because the instruction footprint of server workloads exceeds the private cache capacity. 
Prior works have examined the CPU components or private cache to improve the instruction hit rate.
The large footprint leads to significant cache misses not only in the core and faster-level cache but also in the last-level cache (LLC). 
We observe that even with an advanced branch predictor and instruction prefetching techniques, a considerable amount of instruction accesses descend to the LLC.
However, state-of-the-art LLC designs with elaborate data management overlook handling the instruction misses that precede corresponding data accesses.
Specifically, when an instruction requiring numerous data accesses is missed, the frontend of a CPU should wait for the instruction fetch, regardless of how much data are present in the LLC.

To preserve hot instructions in the LLC, we propose \textbf{Garibaldi}, a novel pairwise instruction-data management scheme. 
Garibaldi tracks the hotness of instruction accesses by coupling it with that of data accesses and adopts management techniques.
On the one hand, this scheme includes a selective protection mechanism that prevents the cache evictions of high-cost instruction cachelines. 
On the other hand, in the case of unprotected instruction line misses, Garibaldi conservatively issues prefetch requests of the paired data lines while handling those misses. 
In our experiments, we evaluate Garibaldi with 16 server workloads on a 40-core machine.
We also implement Garibaldi on top of a modern LLC design, including Mockingjay. 
Garibaldi improves 13.2\% and 6.1\% of CPU performance on baseline LLC design and Mockingjay, respectively. 

\end{abstract}

\makeatletter
\renewcommand{\@titlefont}{\fontsize{17}{21}\selectfont\sffamily\bfseries}

\patchcmd{\@mktitle@iii}
  {\@title\@translatedtitle}
  {\vspace*{-5mm}\@title\@translatedtitle\vspace*{2mm}}
  {}{}

\patchcmd{\@mkauthors@iii}
  {\par\bigskip}
  {\par\vspace*{2mm}\bigskip}
  {}{}

  \renewcommand\nolinkurl[1]{\sffamily{#1}}    


\patchcmd{\KV@ACM@@acmcopyrightmode}
  {%
    \ClassError{\@classname}{%
      Sorry, Creative Commons licenses are\MessageBreak
      currently not used with ACM publications\MessageBreak
      typeset by the authors}{%
      Please use nonacm option or ACM Engage class to enable CC licenses}%
  }
  {}{}{}
  

\makeatother

\maketitle

















\section{Introduction}


Recent server workloads are characterized by significant frontend bottlenecks ~\cite{intro_server_ref_2,intro_server_ref_3,intro_server_ref_4,cloudsuite,emissary}. 
These workloads leverage deeper and more complex software stacks, incorporating extensive kernel operations, virtualization layers, and networking tasks. 
As a result, the instruction footprint in the memory hierarchy has grown substantially and degrades instruction cache performance. 
This footprint increase leads to frequent misses not only in L1 instruction cache~\cite{ispy,fdip,armfdip} but also in L2 unified cache~\cite{emissary}.



In recent years, the architecture community has devoted significant effort to mitigating frontend bottlenecks in servers through advanced branch prediction, instruction prefetching, and partitioning-based instruction protection. 
The techniques are either tightly integrated with the frontend pipeline ~\cite{armfdip,fdip,pdip} or rely on directly monitoring pipeline stall events~\cite{emissary,ispy,intro_server_ref_4}.
However, these studies have a limitation in that they have been developed on a pipeline-centric and per-core basis.
The prior solutions primarily optimize the faster upper-level caches and do not take into account the last-level cache (LLC) that is physically distant from the pipeline and shared among all cores.

\begin{figure}[t] 
    \centering 
    \includegraphics[page=2, width=1.0\linewidth]{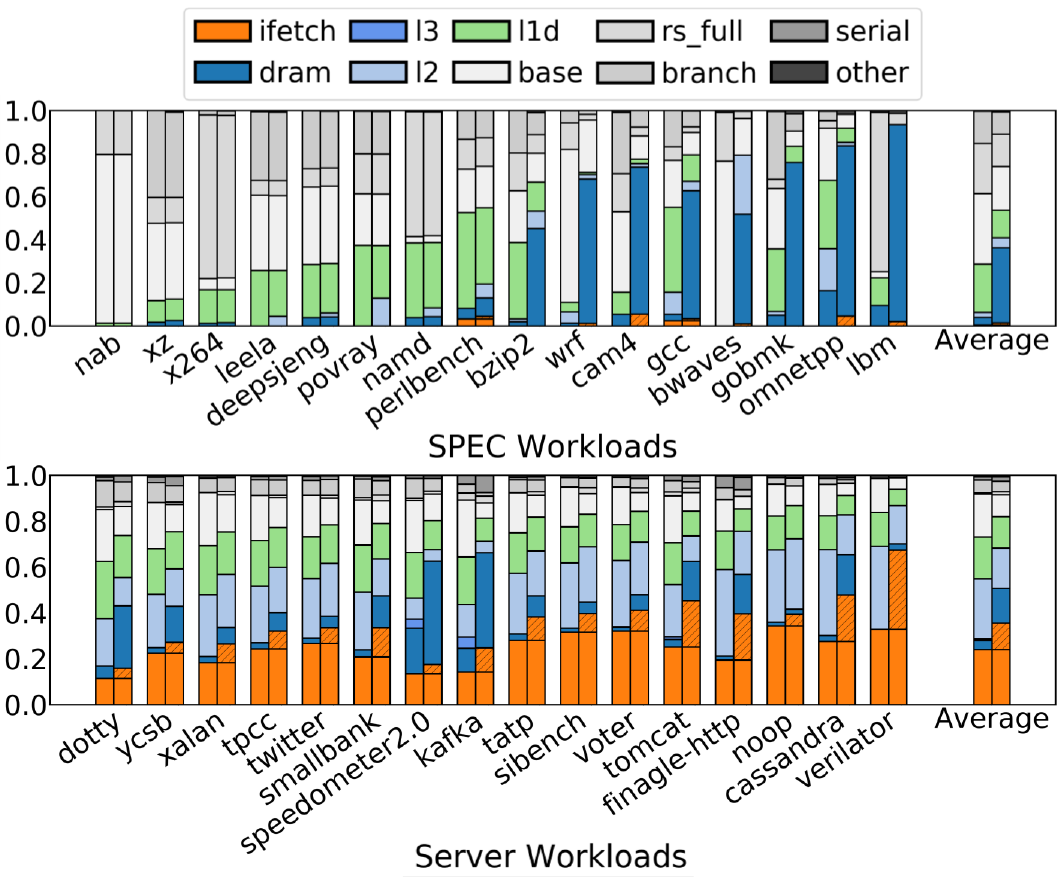}
    \vspace{-7mm}
    \caption{CPI stack difference between core counts of 1 (left bar of each pair) and 40 (right bar of each pair) of spec (top) and server (bottom) workloads.}    
    \label{fig_cpi_stack}
    \vspace{-5mm}
\end{figure}

We observe that server workloads incur significant instruction accesses reaching the LLC even with advanced instruction prefetching~\cite{ispy} and branch prediction~\cite{tage}.
Fig. \ref{fig_cpi_stack} illustrates the normalized CPI stack of SPEC workloads (\cite{spec2017,spec2006}) and server workloads (\cite{dacapo,oltpbench,browserbench,chipyard,renaissance}). We measure the CPI of workloads in the single-core and multi-core (40 cores) configurations with a state-of-the-art LLC management scheme~\cite{mockingjay}.
Interestingly, in the server workloads, the CPI of instruction fetch (ifetch) is one of the largest sources of performance degradation, whereas for SPEC workloads, it is negligible. 
In other words, server workloads experience a substantial frontend bottleneck.
Moreover, this problem is further exacerbated in the multi-core CPU (right bars) compared to the single-core CPU (left bars).
The ifetch bottleneck increases by approximately one-third, primarily due to the LLC contention between multiple cores.


Recent LLC management schemes developed for traditional workloads~\cite{hawkeye,mockingjay} have not considered the instruction footprint in the LLC. 
As these schemes focus on handling massive data demands, the LLC is highly susceptible to instruction cache misses. 
They prioritize caching low-reuse distance data lines (\textit{hot} data) by a practical implementation of Belady’s algorithm.
They take for granted that instruction lines associated with hot data lines are also frequently reused. When the LLC can cover both the instruction and data footprint, the instruction can be cached, especially in traditional workloads.
However, as discussed in Fig. \ref{fig_cpi_stack}, the instruction miss of the LLC is the main bottleneck in the server workloads.
Therefore, the LLC management should evolve to have a mechanism for reducing instruction misses.

In this paper, we analyze how often the instruction and data are reused in the LLC (reuse distance). 
 Our analysis identifies that instruction lines often exhibit long reuse distances ('cold'), while the data lines related to those instruction lines tend to have short reuse distances ('hot'). 
 This imbalance leads to a significant portion of instruction lines being evicted by caching their hot data lines.
 We define this phenomenon as the \textit{instruction victim} problem.


The instruction victim comes from the LLC access pattern of multi-core server workloads.
To understand the unique aspects of server workloads, we conduct several experiments.
In those experiments, we uncover the differences in the instruction/data usage patterns between the SPEC workloads and server workloads.
In the case of SPEC workloads, a few hot instruction lines lead to a number of dispersed data line accesses (\textit{few-to-many}).
On the other hand, the server workloads tend to leverage scattered instruction lines that incur a few hot data line accesses (\textit{many-to-few}). In this \textit{many-to-few} access pattern, cold instructions are frequently evicted in favor of other data cachelines, as their miss cost does not account for the hot data accesses they trigger. This leads to delayed access to cached data lines due to instruction fetch bottlenecks.

When an instruction line is missed in the cache, the CPU suffers from frontend stalls, delaying the triggering of the corresponding data accesses, even if the data lines retrieved by that instruction line are already present in the cache. 
In other words, one instruction miss is much more costly than one data miss.
In addition, the hotter data lines are more likely to be cached, and the cost of the waiting instruction fetch further increases. 
To minimize the cost of cache and improve its overall efficiency, we present a coordinated approach to managing instruction and data cachelines based on their interdependencies. 
We strive to propagate the hotness of data cachelines to the associated instruction cachelines.

To this end, we propose \textit{Garibaldi}
\footnote{Named after \textit{Giuseppe Garibaldi}, the Italian revolutionary leader who united disparate kingdoms into a single, strong nation—mirroring our scheme’s goal of integrating instruction and data management into a cohesive LLC strategy.
}
, a novel instruction-data management scheme that operates orthogonally to existing LLC data management policies.
At the core of Garibaldi is a {\em pair table} that tracks the pair-wise relationship between instruction and data lines in the LLC, propagating the hotness of data lines to their corresponding instruction lines. 
The pair table is designed to integrate seamlessly with existing LLC management schemes, leveraging their strengths in tracking hot data lines. 
For each data access to the LLC, the pair table records \textit{miss\_cost} of the corresponding instruction line.
The \textit{miss\_cost} is incremented or decremented depending on whether the data access is a hit or a miss under the existing LLC management.


With the pair table, we utilize a selective instruction protection policy that prevents the eviction of high-cost instructions ({\em instruction victims}) due to the contention with data lines.
To utilize the recorded hotness of instruction lines in a lightweight manner, we adopt a query-based selection technique~\cite{qbs}, originally proposed to protect inclusion victims in inclusive LLCs. When the LLC management scheme selects an instruction line as a victim, the pair table is queried. If the corresponding \textit{miss\_cost} exceeds a predefined threshold, the instruction line is protected from eviction. Otherwise, it is deemed to be followed by \textit{cold} data accesses, and the instruction line is evicted. 
By leveraging the pair table, Garibaldi additionally supports a data prefetch scheme.
If the cache retrieves an unprotected instruction line that is not present in the cache but recorded in the pair table, Garibaldi makes the cache bring both the instruction line and its paired data lines.

To evaluate Garibaldi, we measure the performance improvement in 16 popular server workloads. 
Garibaldi is compared with state-of-the-art LLC management schemes, including Mockingjay~\cite{mockingjay}, as well as other advanced approaches such as Hawkeye~\cite{hawkeye} and DRRIP~\cite{rrip}.
Since these LLC schemes are orthogonal to Garibaldi, we equip Garibaldi with these LLC designs.
When integrated with Mockingjay, Garibaldi achieves a geometric mean speedup of 13.2\% over LRU, compared to a 6.1\% speedup achieved by Mockingjay alone.

\noindent In summary, our contributions are as follows:
\begin{itemize}[noitemsep, leftmargin=*]

\item We identify the instruction miss problem in the shared LLC for server workloads, a challenge overlooked by both general shared LLC management schemes~\cite{mockingjay,hawkeye,ship,rrip} and server-specific private cache management approaches~\cite{emissary,pdip,fdip,armfdip}.

\item We conduct a detailed analysis of the mechanism behind instruction cacheline misses in the LLC for server workloads. Our observation is that in server workloads, the hotness of data cachelines is not naturally transferred to the corresponding instruction lines, resulting in the \textit{instruction victim} problem.  

\item We propose \textit{Garibaldi}, a novel pairwise instruction-data management scheme for shared LLCs. Garibaldi selectively protects \textit{instruction victims} from eviction and conservatively prefetches associated data lines when unprotected instruction lines are accessed.  

\end{itemize}

\section{Background}

Caching reduces memory latency by bringing instructions and data closer to the pipeline. Fig. \ref{fig:fig formulationsss} illustrates the timeline of instruction and data fetch processes, along with optimization techniques.


\subsection{Cache Optimizations for Instruction Footprint\label{sec:icache_related_work}}
\begin{figure}[t]
    \centering
    \includegraphics[width=1\linewidth]{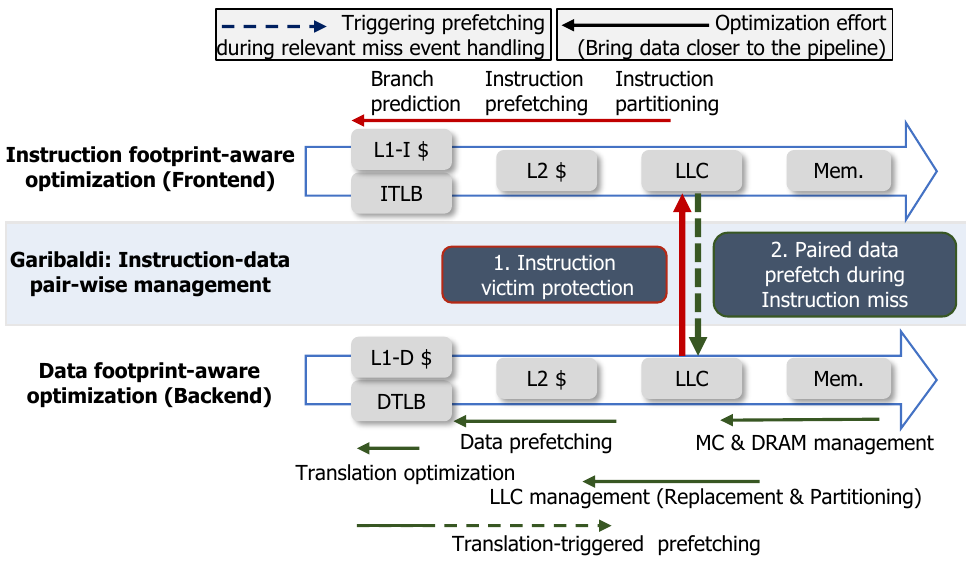}
     \vspace{-6.5mm}
    \caption{Timeline of instruction and data fetch processes, along with optimization techniques.}\label{fig:fig formulationsss}
          \vspace{-1.5mm}
\end{figure}

Instruction caching focuses on mitigating the large instruction footprint in server workloads. Instruction cache misses lead to pipeline stalls, making these techniques tightly coupled with the frontend pipeline, as shown in the upper left of Fig. \ref{fig:fig formulationsss}.

\noindent \textbf{Advanced Branch Predictors:} 
Due to the complex control flow in server workloads, branch predictors must scale to handle a massive instruction footprint~\cite{tage} by effectively caching branch metadata such as branch targets~\cite{thermometer,iandbtbrepl}.

\noindent \textbf{Instruction Prefetching:} 
Another key research area involves prefetching instructions to reduce instruction footprint. Modern server processors often integrate instruction prefetching tightly with the frontend pipeline and branch predictors~\cite{fdip,armfdip,pdip} or leverage profiling data~\cite{ispy,intro_server_ref_4}. Both approaches exploit program execution context to prefetch instructions effectively.

\noindent \textbf{Instruction Partitioning:}
As instruction footprints exceed the capacity of L1 instruction caches, approaches like Emissary~\cite{emissary} target unified L2 caches. Emissary assumes private L2 caches per core and partitions a subset of L2 cache ways to prioritize instruction lines likely to cause decoder stalls. This design uses pipeline event information, such as decoder starvation signals, to guide cache management decisions.


\noindent \textbf{Difficulties of Instruction Cacheline Management in LLC:} 
The LLC, which is shared among all cores, lacks execution context and pipeline stall information except for the program counter (PC). This makes it difficult to attribute LLC performance to individual core’s pipeline, making it hard to measure the cost of instruction misses. Prior approaches~\cite{server_inclusion} addressed this by relaxing inclusion and increasing L2 cache size, shifting instruction line management to private L2 caches—leaving the challenge of managing instruction lines in the LLC as an open problem.

\begin{figure*}[t]
    \centering 
    \includegraphics[page=5, width=0.97\linewidth]{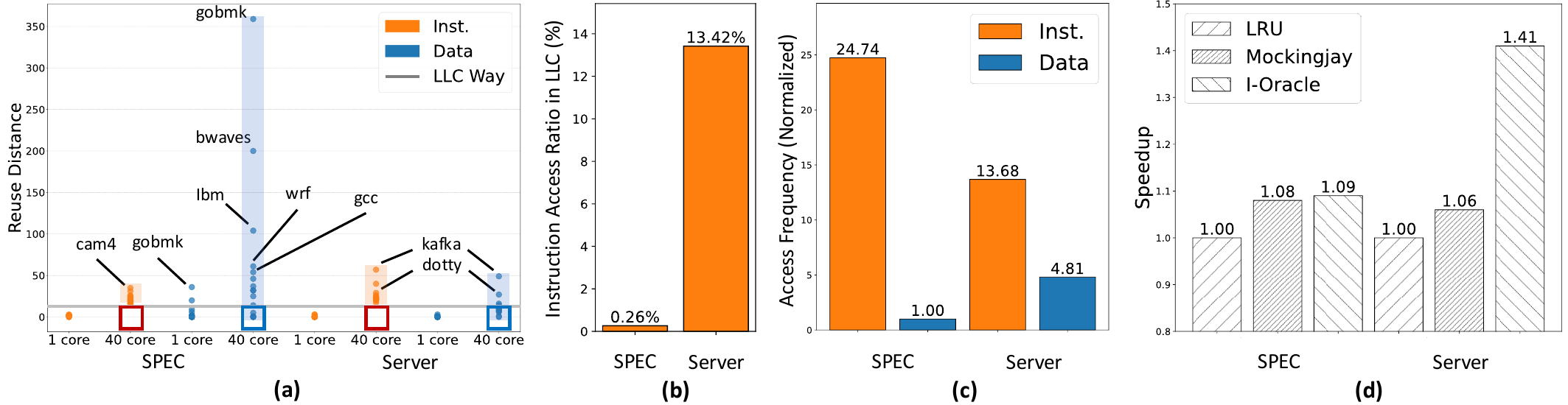}
        \vspace{-2mm}
    \caption{(a) Average reuse distance for each workload, categorized by instruction and data. (b) Average instruction access ratio to total LLC accesses. (c) Average access count per cacheline, categorized by instruction and data. (d) Speedup with state-of-the-art LLC management and potential benefits of instruction management at the LLC.}
    \label{fig_reuse_distance}
    \vspace{-2mm}
\end{figure*}

\subsection{Cache Optimizations for Data Footprint}\label{ssec_data_cache}


Once instructions are fetched and executed in the backend pipeline, they often trigger data accesses. Another focus of cache development is optimizing data access caching for memory-intensive workloads, as shown in the lower part of Fig. \ref{fig:fig formulationsss}.


\noindent \textbf{Translation Optimization:} As a working set of memory-intensive workloads may exceed the range that TLB covers, there becomes a translation bottleneck. Prior work tries to mitigate this by optimizing the translation~\cite{colt,midgard}.

\noindent \textbf{Translation-Triggered Prefetching:} This approach prefetches data during a TLB miss, leveraging the observation that TLB misses often lead to LLC misses~\cite{transtriggered}. Overlapping the latencies of TLB and data misses can reduce overall memory access latency.


\noindent \textbf{Prefetcher:} 
The data access with the translated address now goes down through the cache hierarchy. Here, we can leverage spatial~\cite{spatial1,spatial2,spatial3,spatial4,spatial5}, temporal~\cite{temporal1,temporal2,temporal3,temporal4,temporal5} or both~\cite{ghb} prefetching to reduce the data access latency.

\noindent \textbf{LLC Replacement Policies:}
Cache replacement policies mostly targeted LLC, as it is the largest on-chip storage that can be shared by all cores. The prior replacement policies leverage PC information to predict the reuse distance of data access~\cite{ship,pdp,hawkeye,mockingjay}. State-of-the-art LLC replacement policy, Mockingjay~\cite{mockingjay} successfully combined the PC-based reuse distance predictor and Belady's optimal replacement policy. They evict lines based on the predicted time of reuse and emulate Belady's optimal replacement policy. However, they do not consider the impact of the instruction line on data access. It is because, as shown in Fig. \ref{fig_cpi_stack}, traditional workloads have a negligible instruction footprint on LLC; thus, the instruction line's impact on data access is not considerable. However, we need a different approach for the server workloads, as they have a large instruction footprint on the LLC.

\rvc{my}{BG-2}

\noindent \textbf{LLC Partitioning Policies:} Cache partitioning, particularly way partitioning, is widely used to reduce LLC contention by isolating specific data types, such as network data~\cite{dontforget} or secure data~\cite{catalyst}.\mf{Bg2} However, modern processors with 32–192 cores share LLC with only 12–16 ways of associativity~\cite{8380,7763,wikichip_zen_4,ice_lake_server,xeon_6780e,zen5_wikipedia,xeon_platinum_8480}, creating significant pressure on each set. This imbalance increases conflict misses, severely affecting LLC performance, especially when isolating instruction lines or other data~\cite{mccore,cosmos}.




\subsection{Instruction-Data Pairwise Optimization}
%
We next motivate our approach: instruction–data pairwise management in the LLC.
As shown in the middle side of Fig. \ref{fig:fig formulationsss}, our goal is to bridge LLC data management with instruction-focused optimizations near the frontend pipeline, thus unifying instruction and data caching to improve overall performance.
Specifically, in this work, we overcome the limitations of existing instruction- and data-focused optimizations by introducing instruction and data pairwise management in the LLC. First, we protect the instruction cacheline in LLC by redefining the cost of instruction misses as the opportunity cost incurred by the delayed access to associated hot data lines. Second, by overlapping unprotected instruction miss handling, we also leverage pairwise information to prefetch the associated cold data lines.
\section{Instruction Victims in LLC}
%
This section presents several experiments to investigate the causes of instruction misses in the LLC.
For analysis, we introduce the concept of the \textit{instruction victim}.
We also evaluate the associated costs in terms of the instruction-data pair.
We model the system as described in  §\ref{sec:experimaental_methodology}; unless otherwise specified, we use the Mockingjay~\cite{mockingjay}.

\subsection{Instruction Victim and Cache Access Patterns} \label{subsec:instruction_miss}
\noindent \textbf{Instruction Victims in the Shared LLC:}
Instructions in the LLC are more likely to be victims of contention than data.
%
We define reuse distance as the number of unique accesses to an LLC set between consecutive accesses to the same cacheline. 
It depends on both hardware configuration and application characteristics. 
In multi-core systems, interleaved accesses increase reuse distance compared to a single-core scenario.
It is used for an LLC contention metric~\cite{eeckhout_reuse_distance,rd1,rd2,rd3}: 
If the increased reuse distance exceeds the LLC associativity, the corresponding cacheline is less likely to be retained.

In Fig. \ref{fig_reuse_distance}(a), single-core configurations show that both instructions and data exhibit short reuse distances.
In 40-core configurations, interleaved accesses increase reuse distances for both instructions and data. 
Notably, instructions are more likely to be contention victims than data.
Squared markers indicate reuse distances within the LLC associativity, where cachelines can be retained by modern replacement policies~\cite{rrip,ship,hawkeye,mockingjay}.
The reuse distance points of some data are still placed under the LLC way bar (blue squares), whereas those of all instructions exceed the LLC way (empty red squares).
The observation can be interpreted that instructions are mostly missed in the LLC due to LLC contention.
In our experiments, the instruction miss rate is
98.91\% in SPEC and 95.88\% in server workloads, while the data miss rate is moderate at 67.45\% in SPEC and 42.11\% in server workloads.

\noindent \textbf{LLC Access Pattern Differences in Server Workloads and Their Impact:}
We identify two major factors to understand unique instruction/data access patterns in server workloads.

%
 One is the ratio of instruction and data accesses in LLC (Fig. \ref{fig_reuse_distance}(b)). 
\rv{revb}{B-accessfreq}The other is how many instruction and data requests access each cacheline on average (Fig. \ref{fig_reuse_distance}(c)). 
Fig. \ref{fig_reuse_distance}(b) shows that the instruction accesses come quite often in server workload as 13.42\% whereas the instruction accesses rarely occur in the SPEC workload (0.26\%). 

The server workloads exhibit a high instruction access ratio.
However, they exhibit smaller access counts per cacheline compared to SPEC workloads, as shown in Fig. \ref{fig_reuse_distance}(c).
In the figure, each cacheline serves an average of 24.74 and 13.68 instruction lines in the SPEC and server workloads.
This means there are many cold instruction lines in the server workloads.
On the other hand, the cacheline provides 1.0 and 4.81 data lines, respectively.
%
The frequent data access indicates that the server workloads have a few hot data.
In other words, in the server workloads, the many cold instruction lines lead to accesses to a few hot data lines.
We called it a \textit{many-to-few} access pattern.
On the contrary, the SPEC workloads use a few instructions to utilize many data.
It is called as a \textit{few-to-many} access pattern.
This pattern is also shown in the frequency-weighted reuse distance in Fig. \ref{fig_reuse_distance}(a). 
In 40-core configurations, the server shows a longer reuse distance (orange-shaded bar) than SPEC. 
In contrast, in terms of data reuse distance (blue-shaded bar), the server workloads have lower reuse distances than SPEC. 
Note that in Fig. \ref{fig_reuse_distance}, because many data lines in (c) are rarely accessed, a frequency-weighted average of data reuse (a) would be higher than the per-cacheline average to see the instruction-data access pattern (c).

Considering that instructions can cause frontend pipeline stalls, managing the instruction and data lines equally is not suitable for workloads exhibiting the \textit{many-to-few} access pattern.
Without accounting for this pattern, prior LLC management strategies prioritize caching hot data over instructions and only serve a subset of many instructions.
To evaluate the potential benefit of reducing instruction misses in server workloads with the \textit{many-to-few} pattern, we conducted an Oracle study where instructions always hit in the LLC after their first access.
Fig. \ref{fig_reuse_distance}(d) shows that in \textit{few-to-many} cases, focusing on only caching hot data is enough.
In the SPEC workloads, a state-of-the-art LLC design named Mockingjay achieves a modest 8.4\% performance improvement over LRU, approaching the 9.2\% gain of the instruction oracle (I-oracle).
On the other hand, in \textit{many-to-few} cases of server workloads, Mockingjay achieves 6.3\% performance improvement over LRU, while Mockingjay with I-oracle achieves 42.47\% improvement over LRU.
In other words, there is still significant headroom for an instruction-specific LLC scheme in server workloads.
\begin{figure*}[t]
    \centering 
    \includegraphics[page=8, width=0.98\linewidth]{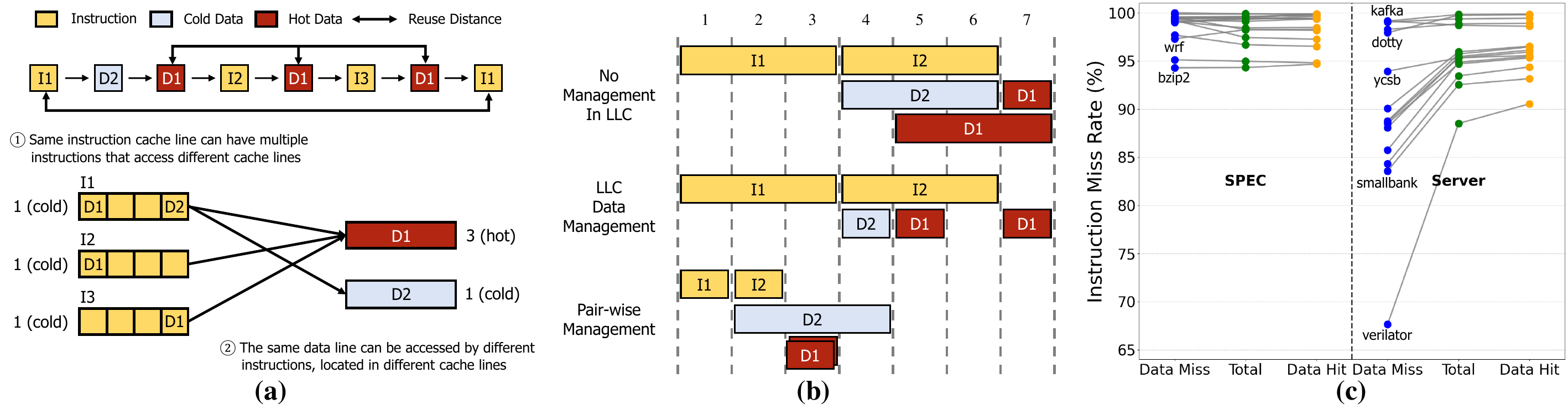}
    \vspace{-2mm}
    \rvc{revb}{B-2}
    \caption{(a) \textit{Many-to-few} access pattern in server workloads. (b) Opportunity cost of instruction misses. (c) \mf{revb} Miss rates of instructions according to the data hotness.}
    \label{fig_Opportunity}
    \vspace{-2mm}
\end{figure*}

\subsection{Many-to-Few Access Pattern and Opportunity Cost of Instruction Miss} \label{subsec:opportunity_cost}
\rvc{my}{moti-4}


We investigate the mechanism of instruction victim problem in server workloads with \textit{many-to-few} access pattern. 
The main reason is that the hotness of the data is not conveyed to the instruction line that triggers this access; that line remains as cold.

Fig. \ref{fig_Opportunity}(a) illustrates the \textit{many-to-few} access pattern of server workloads. 
The cachelines might contain multiple instructions (\ttt{I1}, \ttt{I2}, and \ttt{I3}), and each instruction leads to multiple accesses to data cachelines.
In server applications with substantial data sharing, data cachelines accessed by multiple instructions (\ttt{D1}) are likely to remain in the cache because of their hotness. \rv{revb}{B-datasharing} For example, in \textit{verilator}, 73.7\% of data lines that hit in the LLC were shared by multiple instructions during their insertion-to-eviction lifecycles. 
In contrast, instruction cachelines might be evicted because they are not reused for a 
long time
after the first access (\ttt{I1}) while other instructions and their associated data cachelines are accessed.

This access pattern highlights the problem of instruction cacheline eviction caused by prioritizing hot data caching.
The \textit{instruction victim} problem arises because even hot data accesses are dependent on instruction fetches to be executed.
For example, in a 4-way set associative LLC, hot data \ttt{D1} may remain cached, while subsequent instruction and data accesses evict an instruction cacheline \ttt{I1}.
When \ttt{I1} is needed again, an instruction cache miss occurs, causing frontend stalls and delaying the hot data access \ttt{D1}.
Thus, caching a few more hot data lines can inadvertently slow down instruction fetches and the subsequent execution of hot data accesses.
This observation underscores the need to evaluate instruction and data accesses not as isolated events but as interdependent pairs, incorporating their combined miss costs into cache management strategies.

\noindent \textbf{Opportunity Cost of Instruction Misses:} 
\textit{Instruction victim} problem exhibits varying opportunity costs depending on the hotness of the data cachelines.
\rv{revb}{B-opportunity} Fig. \ref{fig_Opportunity}(b) illustrates the execution timeline when the same set of instruction and data accesses are managed under conventional LLC management schemes and when additional instruction management is applied.
Under conventional LLC management, data cachelines are efficiently cached, resulting in hits.
However, dependency on instruction misses limits performance improvements.
By incorporating instruction management, protecting instructions can lead to performance gains, even if some data cachelines transition from hits to misses (e.g., \ttt{D2}).


The opportunity cost of instruction misses is significantly higher when the associated data is hot.
For example, comparing \ttt{I1} and \ttt{I2} with their first data accesses highlights this difference.
Without cache management, completing the first data access takes 6 cycles for \ttt{I1} when \ttt{D2} misses, compared to 4 cycles for \ttt{I2} when \ttt{D1} hits.
With pairwise management ensuring both \ttt{I1} and \ttt{I2} hit, the time for \ttt{I1} decreases to 4 cycles (1.5x improvement), and \ttt{I2} drops to 2 cycles (2x improvement).
Prioritizing instruction cachelines based on the hotness of their associated data, as seen with \ttt{I2}, significantly reduces critical access delays and improves overall execution efficiency.

If hot data are associated with relatively hot instructions, the \textit{instruction victim} problem may be less significant.
To examine this relationship, we analyze the miss rate of instruction cachelines based on whether their associated data cachelines resulted in hits or misses.
\rvc{my}{moti-5}
\rv{revb}{B-imiss} Surprisingly, Fig. \ref{fig_Opportunity}(c) shows the instruction miss rates associated with hot data accesses are higher than those associated with cold data accesses in almost all server workloads, with the exception of \workname{xalan}.
For instance, in the case of \workname{verilator}, the $Miss Rate_{Data Hit}$ is as high as 90.54\%, whereas the $Miss Rate_{Data Miss}$ is significantly lower at 67.63\%.

By addressing this disparity between opportunity costs and miss rates through pairwise management of instructions and data, we can enhance the performance of server applications, particularly in multi-core environments.

\section{Pairwise Instruction-Data Management in Shared LLC}
In this section, we provide a high-level overview of leveraging paired instruction and data access information to mitigate data access sequence stalls caused by instruction misses in the shared LLC.
First, we define the cost of instruction misses in the LLC within the context of corresponding data cached by modern replacement policies, and introduce a pair table to track this miss cost.
Second, based on this, we propose a method to selectively protect \textit{instruction victims} from eviction caused by competing data accesses.
Finally, we introduce how pair-wise information can further be utilized to reduce unprotected instruction serving delays.  

\begin{figure*}[t]
    \centering
    \includegraphics[page=9,width=1\linewidth]{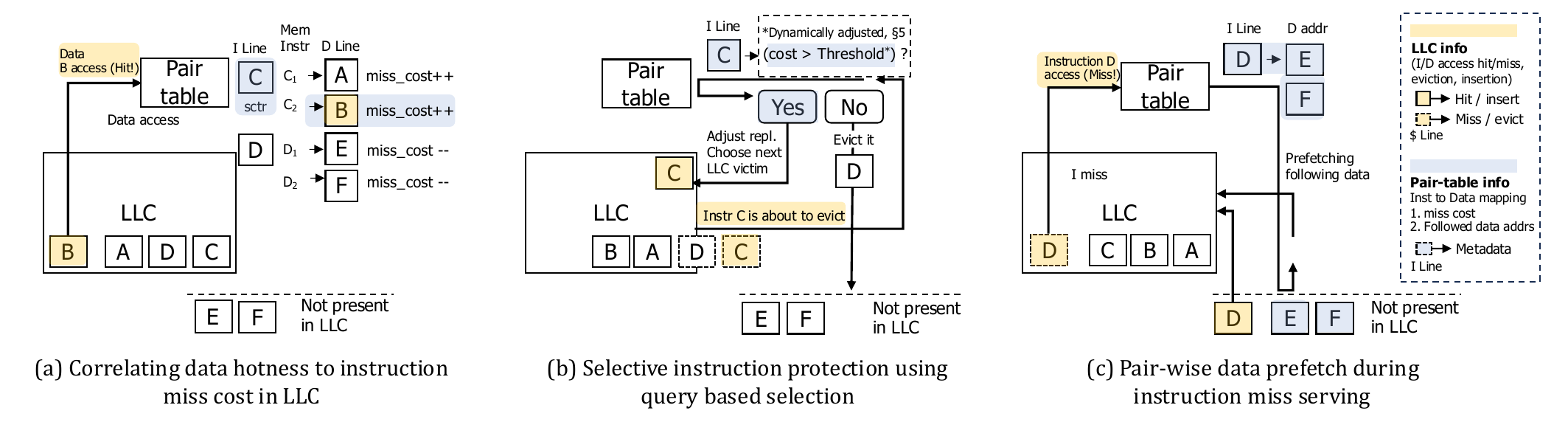}
     \vspace{-7mm} 
    \caption{Instruction-data pair-wise LLC management. (a) Data hotness is paired with the instructions that trigger it. (b) During replacement, queries protect \textit{instructions} associated with hot pairs (\textit{connect back}). (c) When serving an unprotected instruction miss, cold \textit{data} from the pair is prefetched (\textit{connect front}).}\label{fig:fighighlevel}
          \vspace{-2mm}
\end{figure*}

\subsection{Cost of Instruction Misses in the LLC}\label{subsec_miss_cost}

We defined the opportunity cost of instruction cachelines that initiate hot data accesses as the highest.
To straightforwardly quantify the hotness of each access, we leverage modern LLC replacement policies~\cite{hawkeye,mockingjay,ship} that already distinguish and effectively cache hot data accesses by closely approximating optimal replacement strategies. The cache hit status in these policies exhibits 90–99\%~\cite{hawkeye,mockingjay} similarity to the hot data set, defined by those that would be reused with perfect knowledge. 
In our framework, data accesses resulting in cache hits within the LLC are classified as \textit{hot}, while those resulting in misses are deemed \textit{cold}.
This classification enables seamless integration with existing LLC management strategies.
The cost of instruction misses is then correlated with the hit/miss status of the subsequent data accesses they trigger, with misses that lead to data accesses being hits considered more costly.

To implement this correlation, we introduce a pair table that tracks instruction-data pairs and uses this information to quantify the cost of each instruction miss.
We employ an \textit{n}-bit saturating counter for each instruction line in the LLC, which is dynamically adjusted—incremented when an instruction miss leads to a hot data access and decremented otherwise.
For example, as shown in Fig. \ref{fig:fighighlevel}(a), consider instruction line \ttt{C}.
If its associated accesses \ttt{A} and \ttt{B} result in cache hits, we increment the cost counter of \ttt{C} each time.
Conversely, for instruction line \ttt{D}, if its subsequent accesses \ttt{E} and \ttt{F} result in cache misses, we decrement the cost counter of \ttt{D}.

\subsection{Selective Instruction Protection}\label{subsec_case2}

To prevent evicting instructions associated with hot data accesses, 
we revisit insights from inclusion victim avoidance mechanisms.
Jaleel et al.~\cite{qbs} identify the occurrence of inclusion victims as the inability to effectively propagate data hotness from faster cache levels to the inclusive LLC.
Similarly, we argue that current LLC management strategies in server environments suffer from inefficiencies—the failure to convey data hotness within the LLC to the instruction cachelines that access these data results in what we term “instruction victim” scenarios.

Inspired by QBS~\cite{qbs}'s lightweight querying mechanism to mitigate inclusion victims, our approach adopts a query-based mechanism to selectively protect instruction cachelines associated with hot data accesses within the LLC.
We leverage the pair table as a guide for this selective protection mechanism.
During eviction decisions executed by a replacement policy (e.g., Mockingjay) in LLC, if the highest eviction candidate is an instruction cacheline, the pair table is queried using its physical address to retrieve the associated miss cost.
Should the total miss cost exceed a predefined threshold, the eviction priority of the instruction cacheline is reset to the lowest level, thereby preventing its eviction.


For instance, as illustrated in Fig. \ref{fig:fighighlevel}(b), instruction line \ttt{C} is identified as having the highest eviction priority due to infrequent reuse.
When a new cacheline is inserted, \ttt{C} would typically be evicted. 
However, the pair table, which actively tracks and stores the costs associated with prior accesses to \ttt{C}, intervenes in the eviction process.
If the total miss cost of \ttt{C} exceeds the predefined threshold, the pair table prevents \ttt{C} from being evicted by redirecting the eviction process to the next candidate.
In this case, the next candidate, \ttt{D}, is evicted because it has a lower cost.
This strategy avoids the need for partitioning the LLC into separate instruction and data regions, an approach that becomes impractical in modern multi-core systems where LLCs are shared among many cores and designed with low associativity to optimize latency and capacity (§\ref{ssec_data_cache}).  

Additionally, to efficiently distinguish between instruction and data cachelines in both the L2 and LLC, we assign a 1-bit instruction indicator to each cacheline.
This indicator, stored in each L2 cache block with a minimal capacity overhead of 0.2\%, identifies whether the request originates from an instruction line (L1I) or a data line (L1D).
The instruction bit is passed to the LLC upon an L2 miss, enabling accurate differentiation at the LLC level.
This hardware-based approach eliminates the latency and periodic reset issues associated with alternative methods like Bloom filters \cite{bloomfilter}.
By selectively protecting instructions tied to hot data through this query-based mechanism, our scheme ensures valuable cached data remains accessible, enhancing both cache utilization and system performance.


\subsection{Pairwise Prefetch during Instruction Miss Handling}\label{subsec_case4}

In addition to selectively protecting instruction cachelines, we propose a pairwise prefetch mechanism to reduce the latency of serving unprotected instructions.
An instruction miss that is not protected in Garibaldi indirectly indicates that its associated data lines are \textit{cold} and are likely to be missed in the cache.
Leveraging this insight, by prefetching these cold data lines proactively, we can overlap the data miss latency with the ongoing instruction miss handling.

To do so, along with missing costs, the pair table tracks the data line addresses that are associated with the instruction line.
When an instruction miss occurs for an unprotected instruction cacheline, we query the pair table to identify the corresponding cold data.
For instance, as illustrated in Fig. \ref{fig:fighighlevel}(c), instruction line \ttt{D}, which had previously been evicted due to its low miss cost, incurs a miss when accessed again, leading to an ifetch stall.
To mitigate this stall, we prefetch the associated data lines \ttt{E} and \ttt{F}, which are suspected to be cold, into the LLC.
The instruction-data pair information stored in the pair table enables this prefetching.
By prefetching these cold data lines, we ensure that data lines are available when accessed, increasing the cache hit rate and improving overall cache efficiency.


  

\begin{figure*}[t]
  \centering
  \vspace{-3mm}
\includegraphics[page=6,width=.9\linewidth]{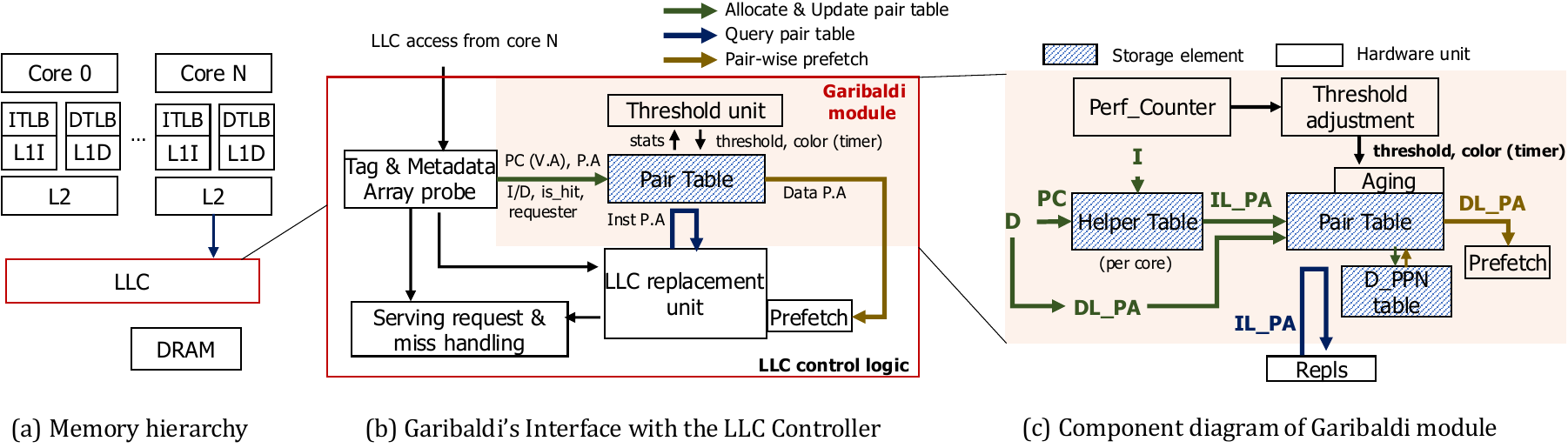}
   \vspace{-3mm}
   
   \caption{Overview of pair table management. (a) and (b) show the physical placement of the Garibaldi module within the memory hierarchy and its interface with the existing LLC controller, respectively. (c) details the Garibaldi module.}
  \label{tb_manage}
        \vspace{-2mm}
\end{figure*}

\vspace{-2mm}
\section{Management of the Pair Table}\label{sec_tb_manage}
%

%
%
Fig. \ref{tb_manage}(a) shows the physical placement of the Garibaldi module within the memory hierarchy, while Fig. \ref{tb_manage}(b) illustrates its interface with the existing LLC controller. Shaded areas indicate the added components in the LLC controller. The three subfigures in Fig. \ref{fig:fighighlevel} correspond to the following actions for the pair table:
\begin{itemize}[noitemsep, leftmargin=*]
    \item \textbf{Allocate \& Update} occurs when the core accesses the LLC. It is the \textit{only operation that writes data into the pair table and uses PC information to track pair} relationships.
    \item \textbf{Query} happens when the LLC replacement policy selects an instruction as a victim, where \textit{managing the threshold and miss cost} of each pair is crucial.
    \item \textbf{Pair-wise Prefetch} is automatically triggered when an instruction miss occurs and there is a matching pair table entry for the missed instruction. Its goal is to \textit{reduce storage overhead} associated with prefetch candidate tracking.
\end{itemize}
  Fig. \ref{tb_manage}(c) details the pair table and its management unit. It illustrates that (1) helper tables track pair relationships (§\ref{ssec_ptrack}), (2) dynamic management adjusts protection thresholds and aged miss costs (§\ref{ssec_dynamic_management}), and (3)%
   decoupled \texttt{D\_PPN} table optimizes storage (§\ref{ssec_dppn}).

\begin{figure}[t]
  \centering
  \includegraphics[width=0.75\linewidth]{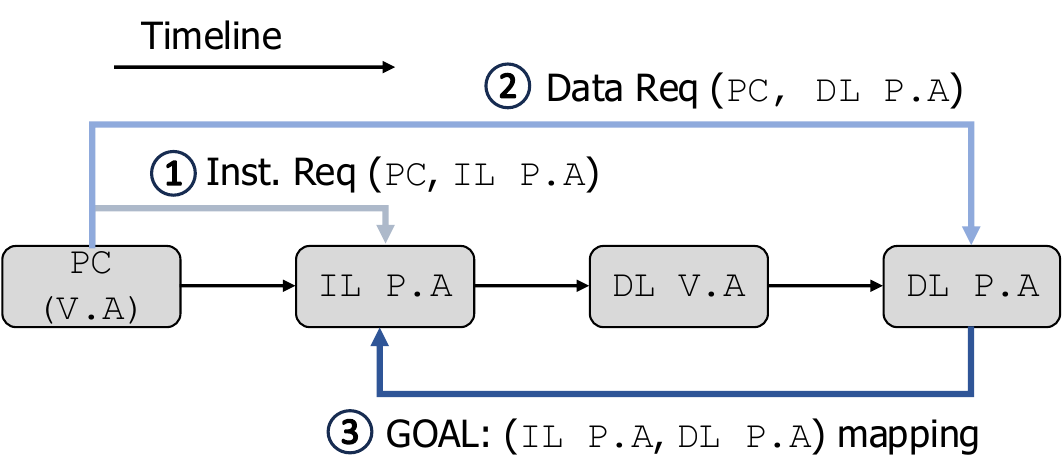}
  \vspace{-2mm}
  \caption{Relationship between instruction and data access.}
  \vspace{-2mm}
  \label{fig:figpc}
\end{figure}


 \subsection{Pair Table Tracking and Allocation}\label{ssec_ptrack}
 Garibaldi manages the LLC by leveraging the mapping between the \textit{physical address} (\ttt{P.A.}) of instruction-data pairs.

\noindent \textbf{Tracking Instruction-Data Pairs in LLC:}
 In Fig. \ref{fig:figpc}, we describe the timeline of tracking instruction and data pairs. 
The key here is \textit{recording} the instruction virtual address (PC) to physical address (P.A.) \textit{translation during the instruction fetch} from LLC. We assume the cache is indexed by the \ttt{P.A.} and that each memory request includes the (\ttt{PC}, \ttt{P.A.}) pair. 
{\textbf{\large \textcircled{\small{1}}}} When instruction fetch, the cache is accessed with a physical address (\ttt{IL\_P.A.}), derived from the PC and aligned to cacheline granularity.  Garibaldi records the PC to IL\_P.A. inside the requester's helper tables in the LLC which is \textit{ITLB-like} structure. 
%
{\textbf{\large \textcircled{\small{2}}}} When the instruction triggers a data access, the physical address of the data line (\ttt{DL\_P.A.}) is obtained. The \textbf{PC of a data access} is the virtual address of the \textit{\textbf{instruction}} that triggers it. Advanced LLC management schemes~\cite{mockingjay,hawkeye,ship} require the PC as a \textit{signature}, since the same physical address (\ttt{DL\_P.A}) may be accessed by different instructions in distinct contexts.
Using the helper table in the LLC, a data access can find the pair instruction line by \textit{\textbf{translating the PC of data access to \ttt{IL\_P.A}} }.
%
%
{\textbf{\large \textcircled{\small{3}}}} After that, the main pair table stores the pair-wise information for the (\ttt{IL\_P.A.}, \ttt{DL\_P.A.}) pair. 
%

\begin{figure}[t]
  \centering
  \vspace{2mm}
  \includegraphics[page=1,width=1.\linewidth]{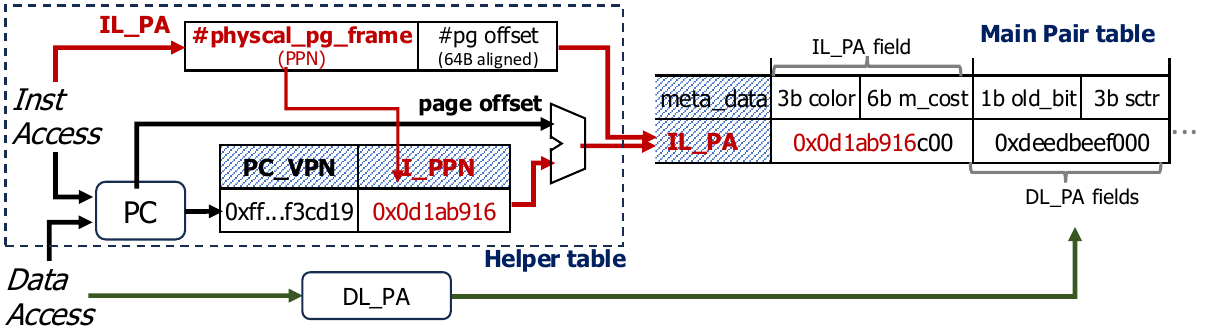}
   \vspace{-6mm}
   \caption{Helper tables record the PC-to-IL\_P.A mapping during instruction accesses, enabling subsequent data accesses to locate corresponding pair table entries with PC.}
 \label{hwtb_manage}
        \vspace{-2mm}
\end{figure}

\noindent \textbf{Pair Table Access and Allocation:}
The helper table and main pair table are allocated/updated during instruction and data accesses to LLC, \textit{respectively}. On the left side of Fig. \ref{hwtb_manage}, the helper table is a set-associative cache that stores \ttt{PC}-to-\ttt{IL\_P.A.} information at page granularity similar to the ITLB. To prevent pipeline instruction fetch interference, we decouple helper tables in LLC from the ITLB—isolating pair table accesses from ITLB operations.

 During the \textit{preceding instruction} access at the LLC, the helper table is allocated/updated to record the page frame mapping—from VPN {\color{orange}0xff…f3cd19} to PPN 0x0d1ab916. With the frame number read from the helper table and the PC’s page offset, \textit{subsequent data} accesses at the LLC can deduce the full IL\_P.A. of their triggering instruction. For example, consider a data access to DL\_P.A. 0xdeedbeef000 with PC {\color{orange}0xff…f3cd19}c00. By combining the PPN from the helper table with the page offset from the PC (0xc00), the IL\_P.A. is deduced as 0x0…d1ab916c00. This approach enables other data accesses to similarly deduce the corresponding instruction addresses on the same page, regardless of whether the instruction is cached in upper-level caches. 
Finally, the main pair table is accessed and allocated/updated with the deduced IL\_P.A. to complete the pair-wise tracking process.

%
%
%

The main pair table is designed to capture the relation between each \ttt{IL\_P.A.} and its corresponding \ttt{DL\_P.A.}
It uses direct-mapped indexing for simple management and minimal lookup latency.
Each entry, as shown in the right part of Fig. \ref{hwtb_manage}, includes \ttt{IL\_P.A.} as a tag, which uniquely identifies the instruction line.
As explained in Section \ref{subsec_miss_cost}, a 6-bit \textit{miss cost} counter tracks the average miss cost of all instructions within a line to prioritize critical instruction lines. Additionally, a \textit{color bits} field supports aging and replacement decisions.  During each allocation/update, the outdated color bits are refreshed.
The table also includes fields for each associated data cacheline, which store \ttt{DL\_P.A.} with metadata monitoring access behavior.
Specifically, each \ttt{DL\_P.A. field} contains a 3-bit saturating counter (sctr) and a 1-bit old bit to track the access status of each field.
%
%
%
%
%
%
%
%
%
%
%
%
\vspace{1mm}
\subsection{Dynamic Management Using Coloring}\label{ssec_dynamic_management}
%
%
Garibaldi dynamically adjusts protection thresholds and miss costs using a synchronized \textit{l}-bit timer. These adjustments are not only critical during queries but also guide allocation and update processes for pair table replacement.
%
%

\begin{figure}[t]
  \centering
  \vspace{2mm}
  \includegraphics[page=1,width=1.\linewidth]{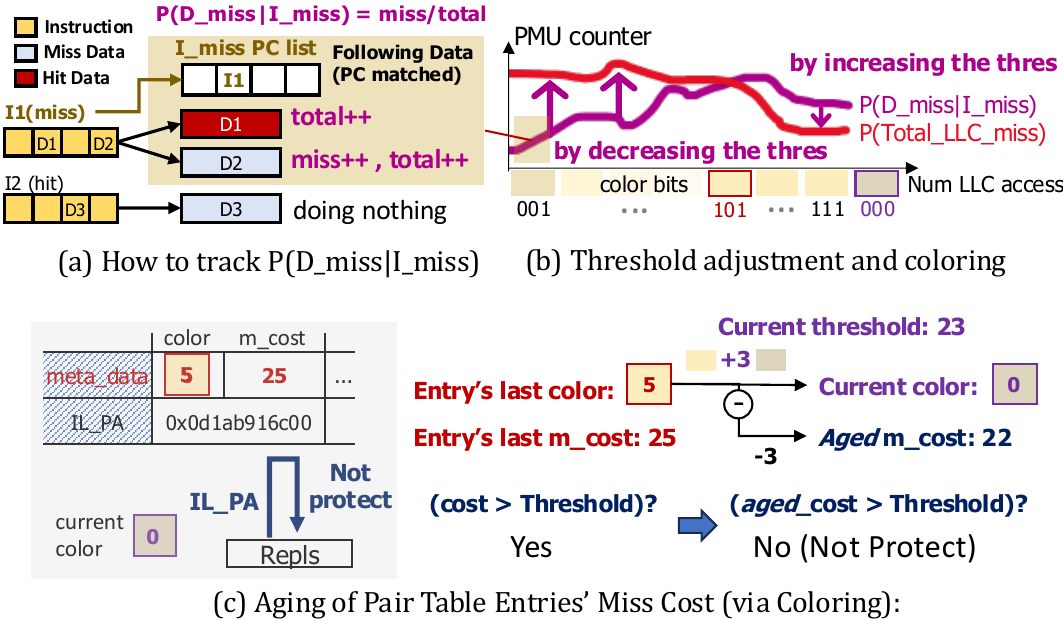}
   \vspace{-5mm}
  \caption{Dynamic adjustment of protection thresholds and miss costs via a synchronized \( l\)-bit timer.}
   \label{hwtb_manage2}
        \vspace{-2mm}
\end{figure}

\noindent \textbf{Dynamic Threshold Adjustment:}
To dynamically adjust the threshold, Garibaldi employs a straightforward approach (Fig. \ref{hwtb_manage2}(b)) in which the cache controller periodically measures the conditional probability of data/instruction miss rates~\( P(D\_miss \mid I\_miss) \). 



  The threshold update unit employs a synchronized \( l\)-bit coloring scheme as a coarse-grained timer for pair-table management. For instance in Fig. \ref{hwtb_manage2}(b), with \( l\) = 3, the timer cycles through 8 color states. Each color represents a period during each \( N \) (e.g., 100K) LLC accesses, and at the end of each period, the threshold is updated using  ~\( P(D\_miss \mid I\_miss) \) and the timer changes to the next color.

   To compute the conditional probability  ~\( P(D\_miss \mid I\_miss) \) —which measures the miss rate for data accesses triggered by instruction misses—Garibaldi employs a dedicated PMU counter that operates as Fig. \ref{hwtb_manage2}(a) during each color (e.g., \ttt{001}):
%
%

\begin{enumerate}[noitemsep, leftmargin=*]
  \item \textbf{Instruction Miss Tracking:} On every instruction miss (\ttt{I1}), the corresponding 64B aligned PC is recorded. A small storage (holding the most recent 10 PCs per thread) is maintained; PCs that are evicted from this storage are automatically excluded from the count.
  \item \textbf{Data Access Matching:} For each data access, if its 64B aligned PC matches one stored in the instruction-miss tracker (\ttt{D1,D2}), both the hit/miss count and the total access count are updated. 
  \end{enumerate}
At the end of each color, by comparing \( P(D\_miss \mid I\_miss) \) to the overall LLC miss rate of that color, the system can adjust the threshold appropriately:
\begin{itemize}[noitemsep, leftmargin=*]
    \item \textbf{Decreasing Threshold:} If \( P(D\_miss \mid I\_miss) \) is significantly lower than the total LLC miss rate, it indicates that many data accesses are successfully served despite instruction misses.
    To prioritize the retention of instruction cachelines, the threshold should be decreased. \ttt{001} of Fig. \ref{hwtb_manage2}(b) is an example of this case.
    \item \textbf{Increasing Threshold:} Conversely, if \( P(D\_miss \mid I\_miss) \) exceeds the total LLC miss rate, it signals that instruction lines are being indiscriminately protected and hurt the total miss rate. 
    To enhance the selectivity of instruction prioritization, the threshold should be increased. 
\end{itemize}
Upon updating the threshold, the system advances to the next color, resets the related PMU counters, and the process repeats.
%
This mechanism ensures a balanced caching of hot data lines and the instructions that access them.


\bigskip 
 \noindent \textbf{Aging of Pair Table Entries’ Miss Cost :}
 In addition to threshold adjustment, we decay the miss cost of pair table entries over time. Without this aging, entries with high miss cost would persist even if their corresponding pairs have not been accessed for an extended period.
 As miss cost is always compared with the threshold, the aging of miss cost should be synchronized with the threshold update. For that purpose, aging process uses the same coloring timer that governs threshold adjustments.


%
During allocate/update of the pair table entry, the current color is recorded for the corresponding entry. 
Suppose an entry has a miss cost of 25 and a color bit of 5, as shown in Fig. \ref{hwtb_manage2}(c).
If IL\_P.A 0x0d1ab916 does not trigger data accesses for a while, it becomes an eviction victim at color 0, which has a threshold of 23.
When the replacement unit queries the IL\_P.A, we check that the entry has a last color of 5 and a miss cost of 25. 
We compare \textit{aged} miss cost with the \textit{current} threshold 23. 
The transition from the last color to the current color passed three steps (5 → 6, 7, 0), resulting in an \textit{aged} miss cost of 22 (i.e., 25 - 3), which is below the threshold of 23 (thus not protected).
Although miss cost is still 25 and higher than the threshold, the aged miss cost makes the IL\_P.A evicted. Note that the entry’s color and miss cost are not updated by the query, remaining 5 and 25, respectively.

\noindent \textbf{Replacement of Pair Table Entries:}
Replacement of pair table entries occurs only during allocation/update, and both accesses and updates employ aging via coloring.
%
In cases of collisions where the indexed entry’s \ttt{IL\_P.A} does not match the accessing \ttt{IL\_P.A1}, and to preserve the high-miss-cost entries in the pair tables, we again use the aged miss cost and threshold comparison as described in the above paragraph.
When aged miss cost is still higher than the current threshold, the entry is preserved. Otherwise, the entry is replaced by the new entry with \ttt{IL\_P.A1}.
The only difference from the query process aging is we update the miss cost with the aged miss cost if the entry is preserved and update the color field of entry to current one.

%



\subsection{DL\_PA Field and Prefetch Management}\label{ssec_dppn}

\begin{figure}[h]
  \vspace{-2mm}
  \centering
  \includegraphics[page=10,width=.98\linewidth]{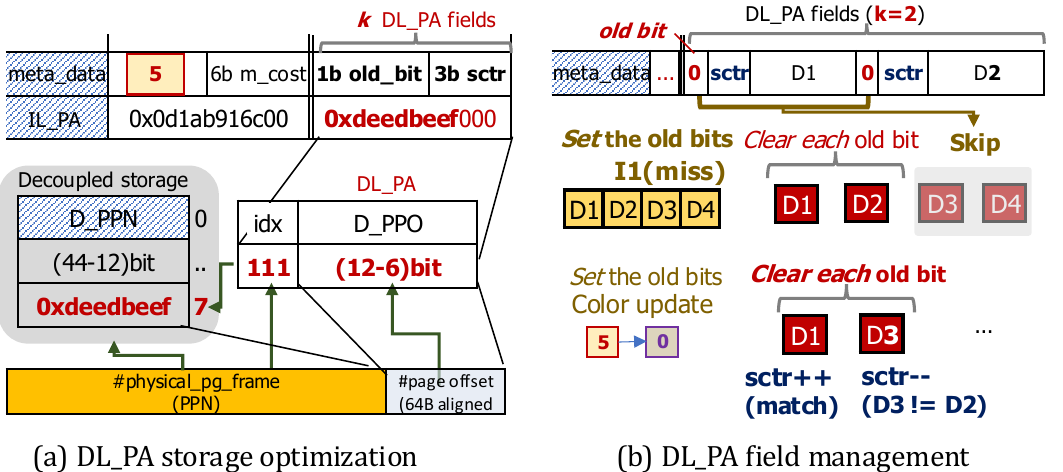}
  \vspace{-1mm}
  \caption{DL\_PA field optimization and management.}
  \label{hwtb_manage3}
        \vspace{-2mm}
\end{figure}

\noindent \textbf{DL\_PA Field Optimization:}
To mitigate storage overhead, each pair table entry stores only a fixed number (\( k \)) of \texttt{DL\_P.A.} fields, as shown in the upper part of Fig. \ref{hwtb_manage3}(a). Furthermore, we leverage the spatial locality among \texttt{DL\_P.A.} fields at page granularity to reduce overhead. Specifically, each data line’s physical address is \textit{split} into the page frame number (\texttt{D\_PPN}) and the offset (\texttt{D\_PFO}). The majority of the address bits (\texttt{D\_PPN}) is stored in a decoupled \texttt{D\_PPN} table and can be shared by many \texttt{DL\_P.A.} fields. 
For example, in a system with 44-bit physical addresses and eight \texttt{D\_PPN} entries, 
each DL\_P.A field retains only the \texttt{D\_PFO} (6 bits) and a 3-bit hashed index (e.g., 111) to access the corresponding \texttt{D\_PPN}.
%
%

\noindent \textbf{DL\_PA Field Management:}
As in Fig. \ref{hwtb_manage3}(b), \textit{each} \texttt{DL\_P.A} field in a pair table entry contains a 1-bit \textit{old bit} indicator to detect the \textit{first} \( k \) data lines following an instruction miss.
This bit is set when either the entry’s color is updated or when an instruction miss (\ttt{I1}) occurs. 
\ttt{DL\_P.A} values (\ttt{D1, D2}) are recorded \textit{only when} at least one of the k fields has \textit{not yet been cleared} of its old bit. Since  k  is typically small (e.g., 2), most data (\ttt{D3, D4}) accesses bypass recording. Clearing the old bit and updating the field are managed by a 3-bit saturating counter (\ttt{sctr}) as shown at the bottom of Fig. \ref{hwtb_manage3}(b):%
\begin{enumerate}[noitemsep, leftmargin=*]
    \item If a \ttt{DL\_P.A} field matches (\ttt{D1}), \textit{increment} its \ttt{sctr} and clear oldbit.
    \item If no match is found (\ttt{D3}), select the \textit{first} field with its old bit set, clear that bit, and \textit{decrement} its \ttt{sctr}.
    \item If the \ttt{sctr} falls \textit{below a threshold} (e.g., 4), replace the field with the new DL\_P.A.
\end{enumerate}
The D\_PPN table also uses a 3-bit sctr similarly for replacement, but without an old bit.

\noindent \textbf{Pair-table Management for Prefetched Cachelines:} 
We assume modern caches distinguish prefetched lines from regular ones. To integrate prefetched lines into the Garibaldi scheme, we query the main pair table using a line’s physical address to determine if it matches an \texttt{IL\_P.A.}. If it does, the line is selectively protected; otherwise, prefetched data lines do not update the pair table, preventing an increase in miss cost.
%
%
Instruction prefetchers generate requests in virtual address space. Whether the engine is profile‑guided \cite{ispy} or an in‑pipeline, branch‑predictor–driven scheme such as FDIP \cite{fdip}, each prefetch follows the normal translation path. As a result, the helper tables still observe the originating \texttt{PC} and record the PC→P.A.\ mapping, allowing prefetched instruction lines to enter pair‑table tracking with no special handling.
\section{Experimental Methodology} \label{sec:experimaental_methodology}

\begin{table}[t]
  \caption{Baseline System Configurations}
  \vspace{-3mm}
  \fontsize{8}{10}\selectfont
  \setlength{\tabcolsep}{4pt}
  \begin{tabular*}{\columnwidth}{@{\extracolsep{\fill}}@{}%
      l                              
      p{0.78\columnwidth}@{}         
  }
    \toprule
    \begin{tabular}[c]{@{}p{0.19\columnwidth}@{}}
      \textbf{Cores}
    \end{tabular}
    &
    \begin{tabular}[c]{@{}p{0.78\columnwidth}@{}}
      40 cores, x86-64, 48KB TAGE/ITTAGE~\cite{tage}, 6-wide issue~\cite{crestmont_anandtech}, OoO, 3.0 GHz, 256-entry ROB, 80-entry LQ, 48-entry SQ, 97-entry IQ, 64-entry ITLB, 48-entry DTLB, 3072-entry STLB~\cite{crestmont_chester}
    \end{tabular}
    \\
    \midrule
    \begin{tabular}[c]{@{}p{0.19\columnwidth}@{}}
      \textbf{L1 I/D caches}
    \end{tabular}
    &
    \begin{tabular}[c]{@{}p{0.78\columnwidth}@{}}
      64KB/32KB~\cite{crestmont_chester}, 8-way, 3-cycle, 10-entry MSHR, I-SPY~\cite{ispy}, Next-line prefetcher (l1d)
    \end{tabular}
    \\
    \midrule
    \begin{tabular}[c]{@{}p{0.19\columnwidth}@{}}
      \textbf{L2 caches}
    \end{tabular}
    &
    \begin{tabular}[c]{@{}p{0.78\columnwidth}@{}}
      4 MB, shared by 4 cores~\cite{sth_sierra}, 16-way, 18-cycle, 64-entry MSHR, GHB prefetcher~\cite{ghb}
    \end{tabular}
    \\
    \midrule
    \begin{tabular}[c]{@{}p{0.19\columnwidth}@{}}
      \textbf{LLC}
    \end{tabular}
    &
    \begin{tabular}[c]{@{}p{0.78\columnwidth}@{}}
      30 MB, shared by 40 cores (0.75 MB/core~\cite{xeon_6780e}), 12-way~\cite{intel_smart_cache}, non-inclusive, 40-cycle, 192-entry MSHR, LRU
    \end{tabular}
    \\
    \midrule
    \begin{tabular}[c]{@{}p{0.19\columnwidth}@{}}
      \textbf{Coherence}
    \end{tabular}
    &
    \begin{tabular}[c]{@{}p{0.78\columnwidth}@{}}
      64 B cache line, MESI protocol
    \end{tabular}
    \\
    \midrule
    \begin{tabular}[c]{@{}p{0.19\columnwidth}@{}}
      \textbf{Memory}
    \end{tabular}
    &
    \begin{tabular}[c]{@{}p{0.78\columnwidth}@{}}
      2-channel DDR5-6400 DRAM, 102.4 GB/s, 49 ns access latency, memory-controller queuing latency modeled
    \end{tabular}
    \\
    \bottomrule
  \end{tabular*}
  \label{tb:baseline}
  \vspace{-4mm}
\end{table}

\noindent \textbf{Modeled System:} We evaluate our design using Sniper~\cite{sniper}, a cycle-level simulator widely used for modeling multi-core x86 architectures. Sniper provides the capability to simulate large core counts with high accuracy and reasonable runtime performance. To capture the complex software stack of server workloads, we employ Gem5~\cite{gem5} in full system mode. This approach generates traces for a single thread of server workloads, encompassing both kernel and user-space activities, which are subsequently processed by Sniper. We ported the Sniper SIFT trace engine to Gem5 source code and added a sanity-checker based on Intel XED. The trace-dumping methodology largely follows the framework outlined in \cite{ms_thesis_sfu}.

For our experiments, we modeled a system resembling Intel’s latest Xeon Sierra Forest processors~\cite{xeon_6780e,sth_sierra,crestmont_chester}, as detailed in Table~\ref{tb:baseline}. This configuration reflects a modern server-grade platform with advanced branch prediction~\cite{tage} and prefetching mechanisms~\cite{ispy}. Specifically, we utilize the TAGE/ITTAGE branch predictor and the I-SPY instruction prefetcher to address the significant instruction footprints of server workloads. Control flow graph (CFG) analysis is integrated into our methodology to model dynamic control flow based on the 80 most recently executed basic blocks. This CFG is leveraged by the I-SPY prefetcher~\cite{ispy} to identify and prefetch instructions likely to incur I-cache misses. Hardware prefetchers for L1d and L2 caches are also included in the baseline configuration. 

To balance simulation runtime with scalability, the system’s core count was reduced from 144 to 40, maintaining an LLC capacity of 0.75MB per core~\cite{xeon_6780e,sth_sierra}. Additionally, the memory subsystem was scaled down from 8-channel to 2-channel DDR5-6400, preserving bandwidth and access latency. The non-inclusive LLC configuration and high-capacity L2 cache are consistent with observed server workloads~\cite{server_inclusion}. We measure the energy using integrated McPAT~\cite{mcpat}.


\begin{table}[t]
  \caption{Storage overheads of Garibaldi}
    \vspace{-2mm}
  \label{tablestorage}
  \resizebox{\columnwidth}{!}{%
  \begin{tabular}{lll}
  \toprule
  \textbf{Structure} & \textbf{Description} & \textbf{Size} \\
  \midrule
  Main pair-table & \begin{tabular}[c]{@{}l@{}}$\bullet$ \#Entries = 16,384 (2\textsuperscript{14}), k=1 \\ $\bullet$  IL\_PA tag (24b) +  miss\_cost (6b) \\ + coloring (3b ) +  valid (1b) = 34 bit \\ $\bullet$ DL\_PA field:   D\_PPO (6b) +  D\_PPN\_idx (13b) \\ +  old\_bit (1b) +  sctr (3b)  = 23 bit \end{tabular} & 120KB \\
  \noalign{\smallskip}
  \midrule
  D\_PPN table & \begin{tabular}[c]{@{}l@{}}$\bullet$ \#Entries = 8,192 (2\textsuperscript{13}), (tagless) \\ $\bullet$ Entry size = D\_PPN (19b) \\ + sctr (3b) + valid (1b) =22 bit \end{tabular} & 32KB \\ 
  \noalign{\smallskip}
  \midrule
  Helper table & \begin{tabular}[c]{@{}l@{}}$\bullet$ \#Entries = 128 (2\textsuperscript{7}) 4-way associativity \\ $\bullet$ Entry size = VPPN (29b) + PPPN (32b) \\ + valid (1b) + sctr (3b) = 64 bit \\\end{tabular} & 1.0KB (per core) \\ 
  \noalign{\smallskip}
  \midrule
  \textbf{Total} &  & \textbf{193.9KB (40 core)} \\
  \bottomrule
  \end{tabular}%
  }
    \vspace{-6mm}
\end{table}

\noindent \textbf{LLC Management Policy:}
We configure Garibaldi as detailed in Table \ref{tablestorage}, assuming a 44-bit physical address (16TB) machine with a four-level page table. The pair table contains \typo{$2\textsuperscript{14}$} entries, keeping storage overhead under 1\% of the baseline LLC capacity. Each instruction line tracks one data line (\textit{k}=1) to minimize the latency of the first data access following an instruction miss, considering the presence of baseline with prefetcher~\cite{ghb}. The helper table includes 128 entries per core, sufficient to cover nearly all accesses to the main pair table. This is based on findings from the 16 server workloads where a 64-entry ITLB achieves 92\%-98\% address translation coverage. Doubling this entry count in the helper table effectively handles address translations. The D\_PPN table, with 8192 entries (half the size of the main pair table), provides a conservative estimate, assuming each instruction line has only one other instruction line accessing the same data page. Overall, Garibaldi’s storage overhead is 193.9 KB for 40 cores, accounting for 0.6\% of the baseline LLC capacity. An additional 1-bit instruction indicator per cache block increases the total overhead to 0.8\% of the LLC capacity.
  Our solution leverages QBS~\cite{qbs}, which uses two parameters affecting LLC operations. \rv{revd}{D-3} First, based on CACTI7~\cite{cacti7} (22 nm), the pair table access latency is 0.331 ns, so we set \texttt{QBS\_LOOKUP\_COST} to 1 cycle. \rv{reve}{E-4} Second, we limit the number of pair table queries per eviction by setting \texttt{QBS\_MAX\_ATTEMPTS} to 2, capping the worst-case eviction overhead at 2 cycles.

Garibaldi is orthogonal to existing LLC management policies and integrates seamlessly with state-of-the-art replacement policies. To evaluate its effectiveness, we implemented Garibaldi on top of Hawkeye~\cite{hawkeye}, Mockingjay~\cite{mockingjay}, and DRRIP~\cite{rrip}. Mockingjay, in particular, mimics the behavior of Belady’s optimal policy. Each policy uses 5-bit ETR/RRPV counters, with history lengths set to $8\times$ associativity size for sampled sets. A total of 4096 sets are sampled to accommodate the 40-core configuration.

  \begin{table}[t]
    \centering
    \caption{Benchmarks used to evaluate Garibaldi}
    \label{tab:benchmarks}
    \vspace{-2mm}
  
    \setlength{\tabcolsep}{1.5pt}
    \fontsize{8}{10}\selectfont
  
    \begin{tabular}{@{}l@{\hspace{10pt}}p{0.69\linewidth}@{}}
      \toprule
      \textbf{Benchmark Suite} & \textbf{Benchmarks} \\ 
      \midrule
      DaCapo~\cite{dacapo} &
        \ttt{cassandra}~\cite{cassandra}, \ttt{tomcat}~\cite{tomcat},
        \ttt{kafka}~\cite{kafka}, \ttt{xalan} \\
  
      Renaissance~\cite{renaissance} &
        \ttt{finagle-http}~\cite{finaglehttp}, \ttt{dotty}~\cite{dotty} \\
  
      \multirow{2}{*}{\makecell[l]{OLTP Bench~\cite{oltpbench}\\(PostgreSQL~\cite{postgresql})}} &
        \ttt{tpcc}~\cite{tpcc}, \ttt{ycsb}~\cite{ycsb}, \ttt{twitter},
        \ttt{voter}, \ttt{smallbank}, \\ 
      & \ttt{tatp}, \ttt{sibench}, \ttt{noop} \\
  
      Chipyard~\cite{chipyard} &
        \ttt{verilator}~\cite{verilator} \\
  
      Browser Bench~\cite{browserbench} &
        \ttt{speedometer2.0}~\cite{speedometer} \\
      \bottomrule
    \end{tabular}
    \vspace{-4mm}
  \end{table}

  \noindent \textbf{Workloads:}
%
%
We choose 16 popular server workloads with diverse characteristics, including database, web server, and in-memory analytics applications, which are all front-end heavy benchmarks~\cite{pdip}. The workloads are selected from the DaCapo~\cite{dacapo}, Renaissance~\cite{renaissance}, OLTP Bench~\cite{oltpbench}, Chipyard~\cite{chipyard}, and Browser Bench~\cite{browserbench} benchmark suites. All of which the detailed list of workloads is provided in Table~\ref{tab:benchmarks}. These applications are also widely used as a representative of modern server workloads in prior studies~\cite{pdip,emissary}.

In the trace dump stage from gem5 full-system simulations, we sampled 100M instructions after a warming period of 10M instructions. When running each trace on Sniper, we first warm up using the first 20M instructions, followed by a detailed simulation for the remaining 80M instructions. Since we use the 40 cores, a total of 3200M instructions are fed to the detailed model of Sniper in each run. For homogeneous workloads, we use the harmonic mean of IPC as a metric. For heterogeneous workloads, we use the weighted speedup metric (=$\sum \frac{\text{IPC}_{\text{shared}}}{\text{IPC}_{\text{single}}}$). Workloads are replayed until every trace has been completed at least one time.

  



\section{Evaluation Results}\label{evalsec}
\subsection{End-to-End Performance Comparison}
\begin{figure}[ht]
    \vspace{-3mm}
    \centering
    \includegraphics[width=1\columnwidth]{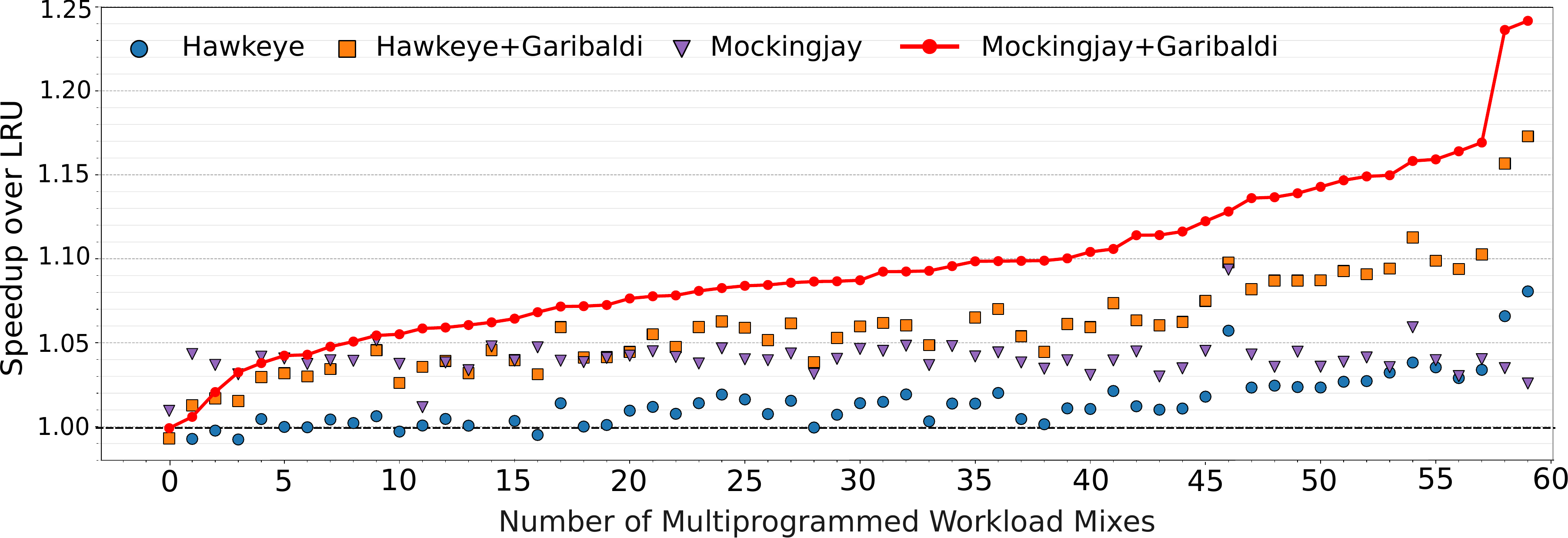}
    \vspace{-6mm}
  
    \caption{End-to-end performance comparison between Garibaldi and state-of-the-art LLC management policies.}
    \vspace{-3mm}
    \label{fig:rlres2}
\end{figure}


We first conduct an end-to-end performance comparison between state-of-the-art reuse distance-based LLC management policies and Garibaldi, built upon these policies. Fig. \ref{fig:rlres2} shows the performance distribution across 60 randomly chosen traces from Table \ref{tab:benchmarks}. The weighted speedup of each workload is normalized to the baseline using LRU, and the workload is sorted by the relative performance of Mockingjay with Garibaldi over LRU.

Hawkeye and Mockingjay, while effective at caching hot data, show limited performance gains in server environments due to the “instruction victim” phenomenon. Specifically, Hawkeye achieves only a geometric mean speedup of 1.3\% compared to LRU. Mockingjay, a state-of-the-art reuse distance-based policy, achieves a slightly higher speedup of 4.0\% over LRU by leveraging more refined reuse distance predictions and prefetch-aware optimizations. 
 While Mockingjay outperforms Hawkeye by providing better data caching, both remain constrained by the detrimental effects of instruction victimization.
In contrast, Garibaldi mitigates the instruction victim phenomenon, unlocking the potential for more timely reuse of hot data. When paired with Hawkeye, Garibaldi achieves a geometric mean speedup of 5.6\%, even outperforming Mockingjay without Garibaldi. When combined with Mockingjay, Garibaldi further improves performance, achieving a geometric mean speedup of 9.3\%. 

The benefits of Garibaldi become even more pronounced in workloads where instruction victimization is prevalent (speedup up to 24\%). Under such conditions, the performance gap between Mockingjay and Hawkeye diminishes, as Mockingjay’s advantage in better data caching is offset by delays caused by instruction victimization. Conversely, in workloads with minimal instruction victimization (e.g., workloads 0–5), Garibaldi’s instruction protection mechanisms introduce a tradeoff, slightly reducing data caching efficiency. As a result, Mockingjay emerges as the most performant policy in these cases. However, for the majority of workloads, the benefits of Garibaldi’s instruction protection outweigh the cost of marginally increased data misses, delivering significant overall performance improvements.

\begin{figure*}[ht]
    \includegraphics[width=0.88\linewidth]{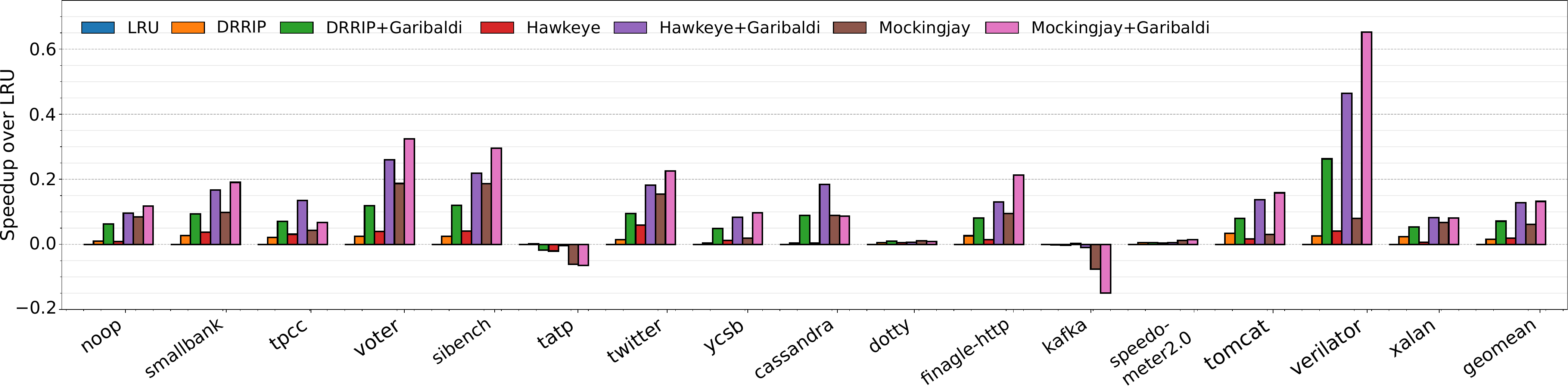}

    \vspace{-2mm}
    \caption{Speedup comparison of Baseline, DRRIP, Hawkeye, mockingjay and Garibaldi under homogeneous workload }
    \label{fig:throughput}
\end{figure*}

\begin{figure*}[ht]
    \vspace{-1mm}
    \includegraphics[width=0.88\linewidth]{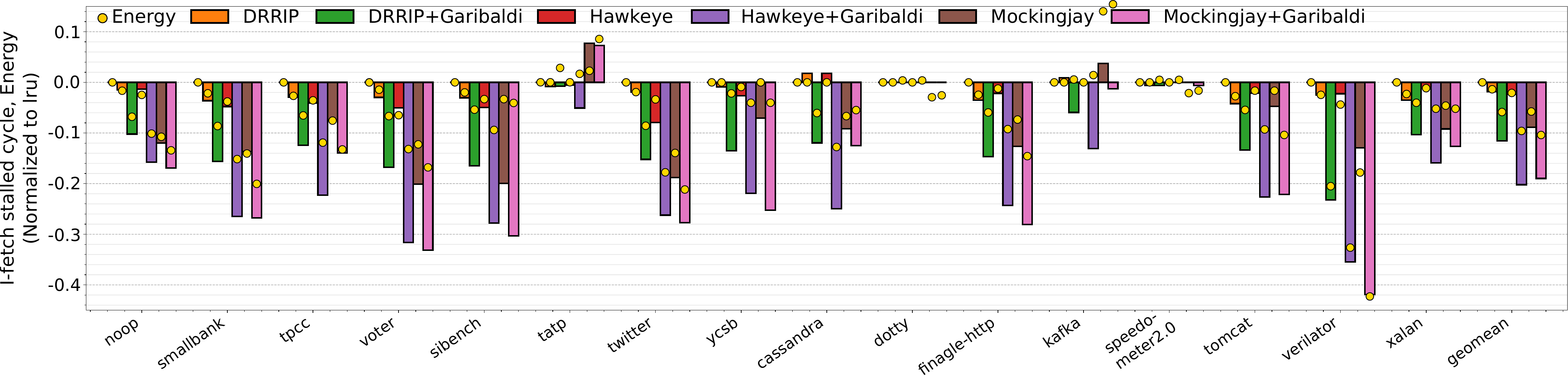}

    \vspace{-3.5mm}
    \caption{Ifetch stalled cycles and energy normalized to LRU} 
\vspace{-3mm}
    \label{fig:dse1}
\end{figure*}
%
\subsection{Per Workload Analysis of Garibaldi}

\noindent \textbf{Performance Analysis:}
Fig. \ref{fig:throughput} presents the speedup of DRRIP, Hawkeye, and Mockingjay with and without Garibaldi under 16 server workloads over baseline with LRU. The results show that Garibaldi easily outperforms the other three policies. 
First, Garibaldi's effect can be seen in all of the policies. 
On average, Garibaldi enhances DRRIP, Hawkeye, and Mockingjay to achieve speedups of 7.1\%, 12.8\%, and 13.2\% over LRU, compared to their standalone speedups of 1.5\%, 1.9\%, and 6.1\%, respectively.
Garibaldi's benefit outweighs the better data caching.
For example, even DRRIP with Garibaldi yields a performance similar to that of Mockingjay without Garibaldi.
But at the same time, the performance after the combined policies abilities ironically determines Garibaldi to cache hot data; performance after Garibaldi has the same order of performance without the Garibaldi.
However, the performance improvement per each policy is different, and not always better data caching policy gets more additional benefits from the Garibaldi.
For example, Hawkeye's benefit from Garibaldi is 10.7\%, while Mockingjay's benefit from Garibaldi is 6.7\%, and DRRIP is 5.4\%. %

The reason for this performance gain can be explained by the ifetch stall cycle.
Fig. \ref{fig:dse1} shows the average ifetch stalled cycle normalized by the LRU.
We can see that the workloads that Garibaldi achieves high-performance gain have a reduction in the ifetch stalled cycle.
For example, \ttt{verilator} where Garibaldi with Mockingjay achieves 65.2\% speedup over LRU, where it reduces 42\% ifetch cycles of LRU.
We can see that the performance benefit from Garibaldi is not uniform across the workloads, from -18\% to 65\% speedup over LRU, which can explain the variance in ifetch stall reduction across the workloads.
The general trend shows that the higher it saves, the ifetch reduction, the better performance gain it gets.
On the geometric mean, whereas Mockingjay alone has a 9\% reduction in the ifetch stall compared to the baseline, the Mockingjay with Garibaldi has an 18\% reduction in the ifetch stall cycle. 
Even when Garibaldi underperforms the baseline (e.g., in \texttt{kafka}), it still reduces ifetch stalls. In \texttt{kafka}, both data and instructions are cold; moreover, the data reuse distance is not significantly lower than that of the corresponding instruction, and both reuse distances shows the longest among all workloads. In such cases of insufficient data caching, protecting instructions with Garibaldi is not beneficial, as it trades off data caching to handle instruction victims.
%


Garibaldi also improves energy efficiency. yellow scatter in Fig. \ref{fig:dse1} shows the energy consumption normalized by the LRU. On average, Garibaldi with Mockingjay saves 10.4\% of energy compared to LRU, which is 5.0\% lower compared to Mockingjay alone. The other policies also show a similar trend. 
The results show that the energy reduction trend is aligned with the ifetch cycle reduction trend except for two workloads, \ttt{kafka} and \ttt{tatp}. This result shows that the removal of unnecessary stalls caused by instruction victimization can reduce energy consumption.
In two workloads (kafka and tatp) the benefits are offset by increased data misses and additional pair table energy, resulting in up to 14.5\% more energy consumption. By contrast, in \texttt{verilator}, Garibaldi reduces energy consumption by 42.3\% by eliminating avoidable ifetch stalls.

\begin{figure}[!h]
    \centering
    \vspace{-2mm}
    \captionsetup[subfigure]{font=small, labelfont=small}
    \begin{subfigure}[t]{0.48\linewidth}
        \centering
        \includegraphics[width=0.9\linewidth]{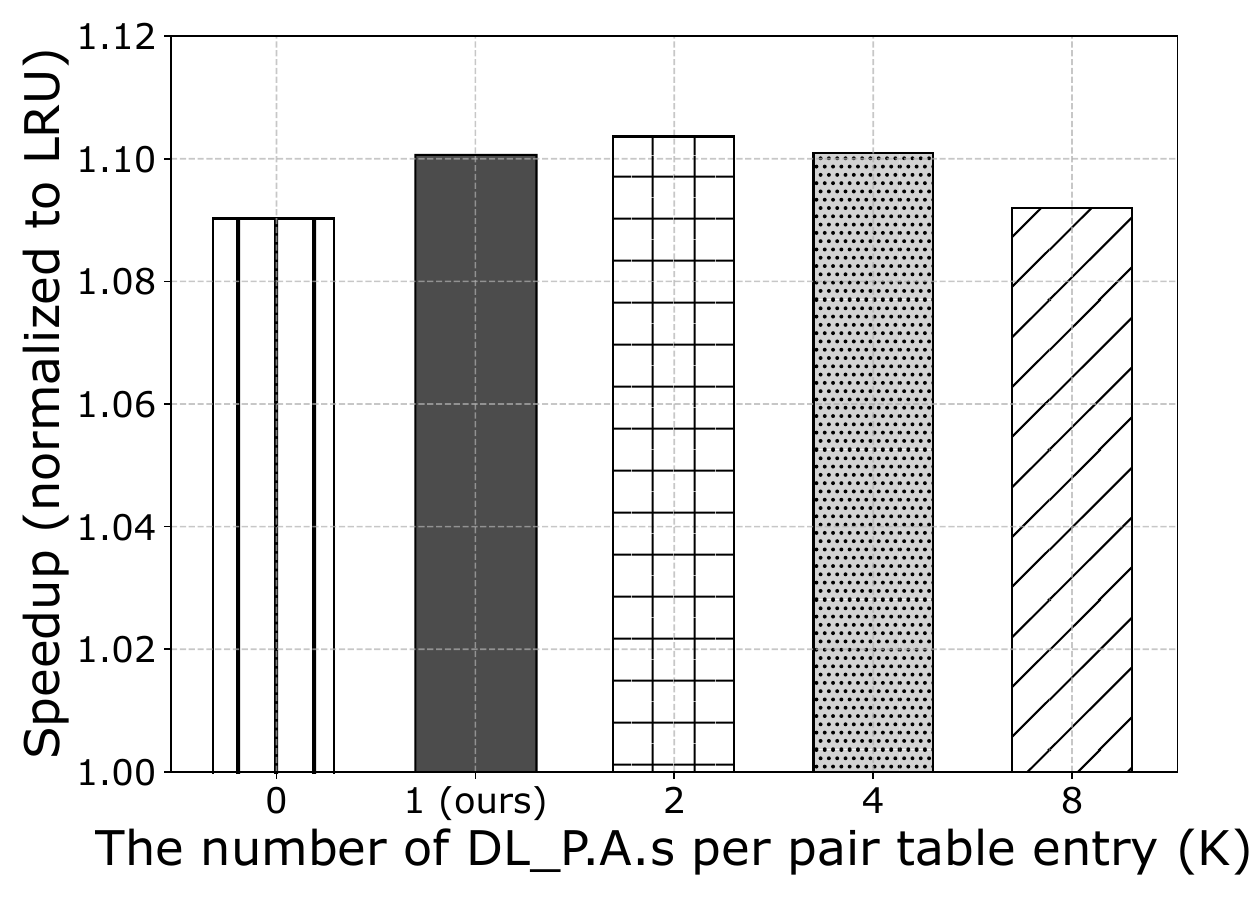}
        \vspace{-2mm}
      {\footnotesize  \caption{\textit{k} data lines per pair table entry}}
        \label{fig:k_sweep}
    \end{subfigure}
    \hspace{-0.3mm}
    \begin{subfigure}[t]{0.48\linewidth}
        \centering
        \includegraphics[width=0.9\linewidth]{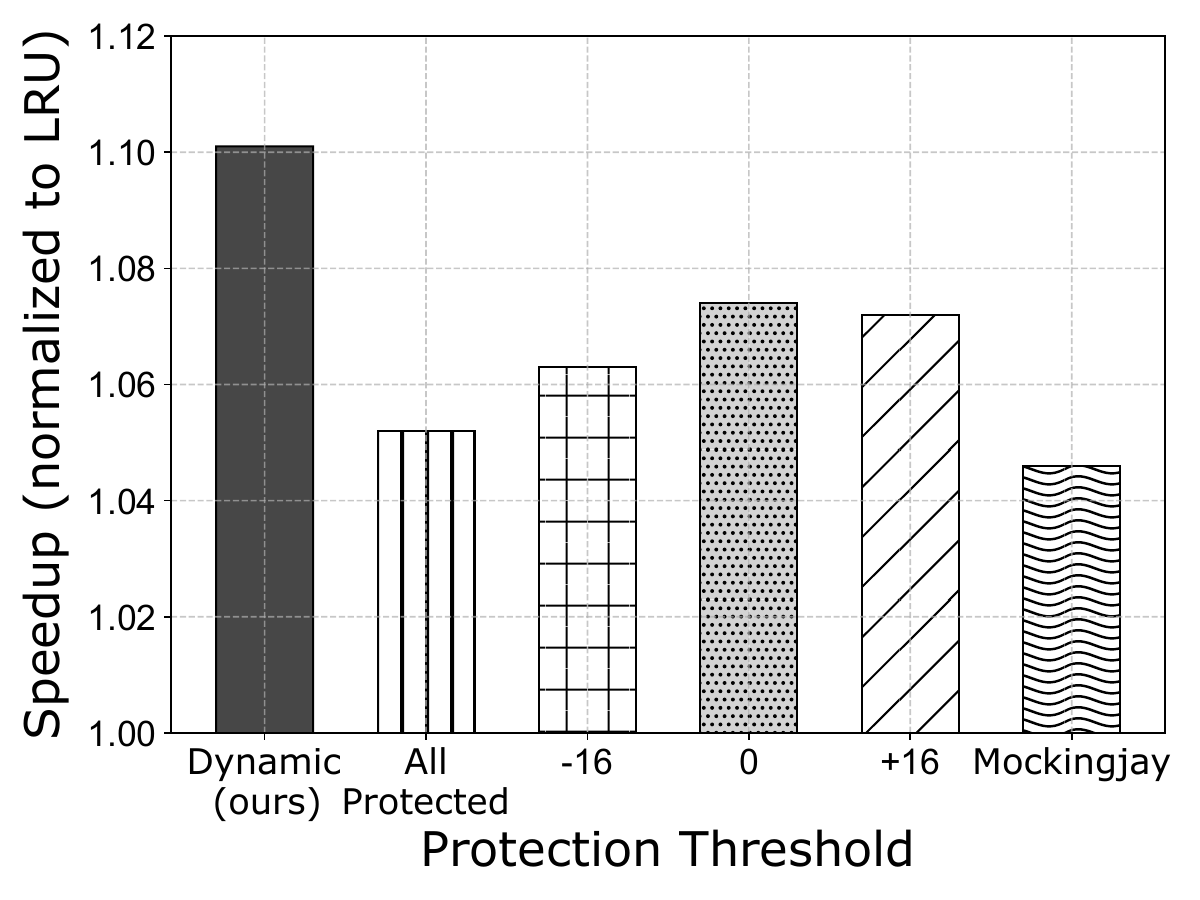}
        \vspace{-2mm}
        {\footnotesize \caption{Protection thresholds as deltas from the init value (\texttt{32}).}}
        \label{fig:protection_threshold_sweep}
    \end{subfigure}

    \begin{subfigure}[t]{0.48\linewidth}
        \centering
        \includegraphics[width=0.9\linewidth]{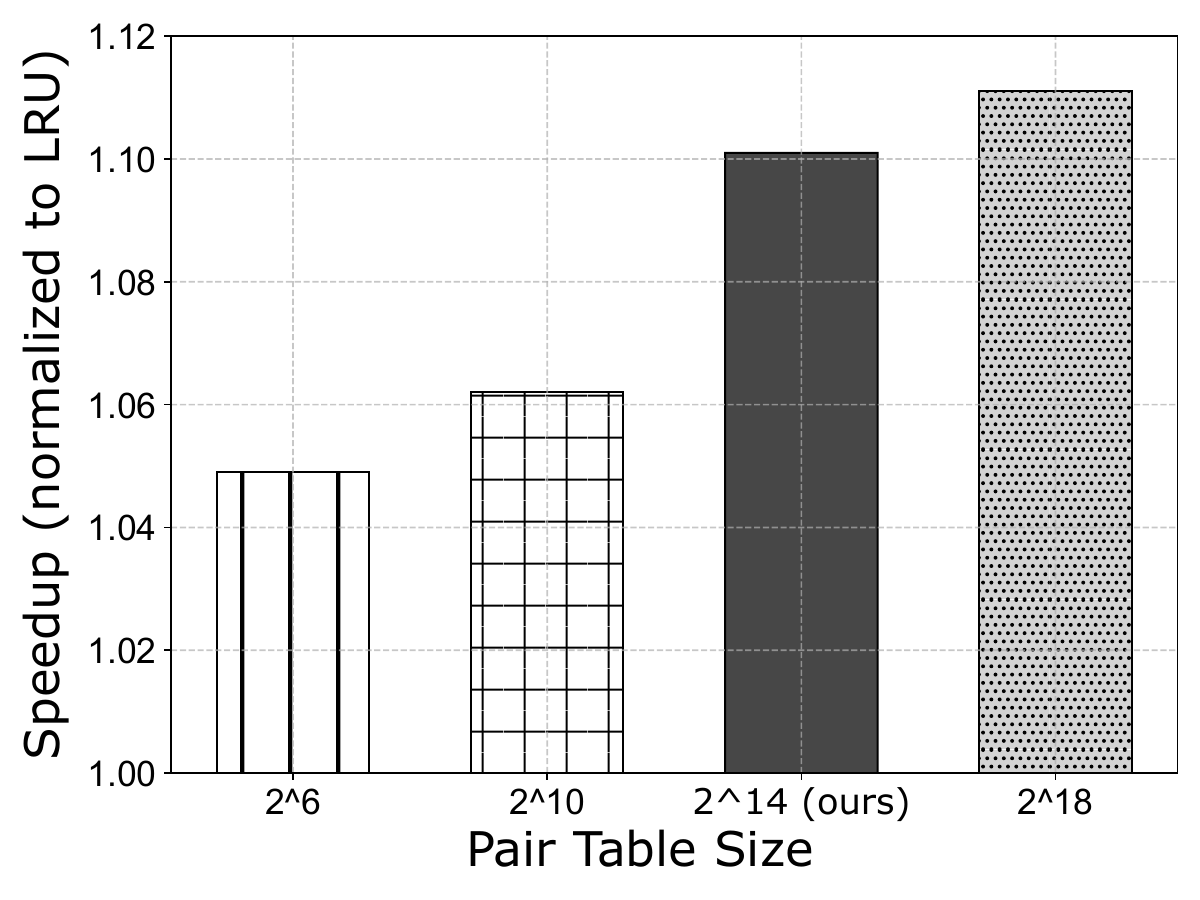}
        \vspace{-2mm}
        {\footnotesize  \caption{Number of pair table entries}}
        \label{fig:pair_table_sweep}
    \end{subfigure}
    \hspace{-0.3mm}
    \begin{subfigure}[t]{0.48\linewidth}
        \centering
        \includegraphics[width=0.9\linewidth]{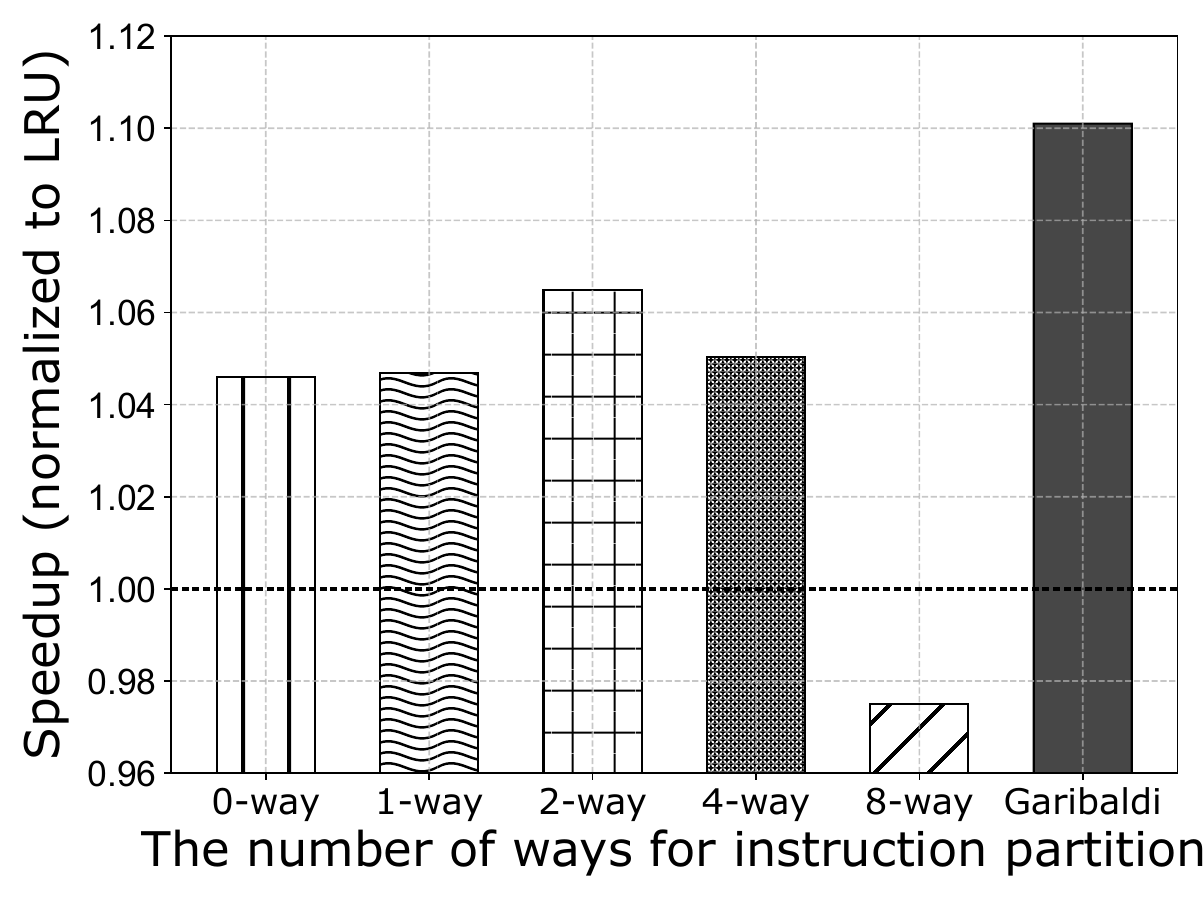}
        \vspace{-2mm}
        {\footnotesize  \caption{Partitioning vs Ours}}
        \label{fig:average_partition_speedup}
    \end{subfigure}
    \vspace{-4mm}
    \caption{Evaluation results showing speedups under different configurations.}
    \vspace{-5mm}
    \label{fig:design_space}
\end{figure}

\subsection{Garibaldi Configuration Analysis}
\label{subsec:desing_space}





We conducted sensitivity studies on 30 workload mixes to evaluate factors affecting Garibaldi’s performance combined with Mockingjay. Fig. \ref{fig:design_space} summarizes the results.

    \noindent \textbf{Number of Data Lines per Pair Table Entry (\textit{k}):} Fig. \ref{fig:design_space}(a) shows the effect of varying the number of data line addresses (\textit{k}) per pair table entry. With no data line address stored (\textit{k} = 0), the geometric mean improvement over LRU is 8.9\%. Small \textit{k} values (\textit{k} = 1 or \textit{k} = 2) perform best, achieving 10.1\% and 10.2\% improvements, as the data prefetcher~\cite{ghb} effectively overlaps subsequent data misses and only needs to address the first one or two misses following an instruction miss. However, prefetching too many data lines using pairwise information is not effective; for instance, \textit{k} = 8 yields a reduced 9.2\% improvement over LRU.

    \noindent \textbf{Protection Threshold:} Fig. \ref{fig:design_space}(b) illustrates the impact of the protection threshold. Garibaldi with a dynamic threshold is compared to fixed-threshold approaches. A zero threshold protects all instruction lines in the pair table, achieving a 5.2\% speedup, while Mockingjay, with no protection, achieves 4.6\%. Selective protection with fixed thresholds performs better: higher thresholds (e.g., \textit{+16}) and lower thresholds (e.g., \textit{-16}) provide 7.1\% and 6.3\% speedups, respectively. The best-fixed threshold (\textit{+0}) achieves 7.4\%, while Garibaldi’s dynamic threshold adapts to workloads, achieving a 10.1\% speedup.
    
    \noindent \textbf{Number of Pair Table Entries:} Fig. \ref{fig:design_space}(c) illustrates the performance impact of pair table size. Since instruction and data line pairs are rarely mapped to the same sets, the pairwise relationship cannot be sampled on a set basis~\cite{qureshi2007adaptive}, necessitating a large pair table size. Larger tables yield better performance, with \typo{$2\textsuperscript{18}$} entries achieving the highest improvement (11.1\% over LRU) but incurring impractical storage overhead (>6\% of LLC capacity). Smaller tables reduce performance: \typo{$2\textsuperscript{10}$} entries provide a 6.2\% improvement, significantly lower than the default \typo{$2\textsuperscript{14}$} configuration (10.1\%), while \typo{$2\textsuperscript{6}$} entries degrade performance to a 4.9\% improvement over LRU.
   


\noindent \textbf{Comparison with Partitioning-Based Instruction Protections:} \rv{reva}{A-2}\rv{revc}{C-6} \rvc{revc}{C-11}
~Fig. \ref{fig:design_space}(d) compares Garibaldi with a partitioning-based approach. Although unrealistic in the LLC (\S\ref{sec:icache_related_work}), we also employ pipeline events~\cite{emissary} to protect only critical instructions (up to \( N \)), mitigating the inefficiencies of naive partitioning. In its best configuration—allocating two ways (\( N \)=2)—it yields a 6.49\% improvement over LRU. However, allocating more than two ways for instructions reduces the cache associativity available to data (\S\ref{ssec_data_cache}), which can lead to performance degradation below LRU levels (as shown for 8-way in Fig. \ref{fig:design_space}(d)). In contrast, Garibaldi’s query-based selection effectively protects instruction lines without reducing cache associativity, delivering superior performance.

\begin{figure}[h]
    \centering
    \vspace{-4.5mm}
    \captionsetup[subfigure]{font=small, labelfont=small}
    \begin{subfigure}[t]{0.48\linewidth}
        \centering
        \includegraphics[width=0.9\linewidth]{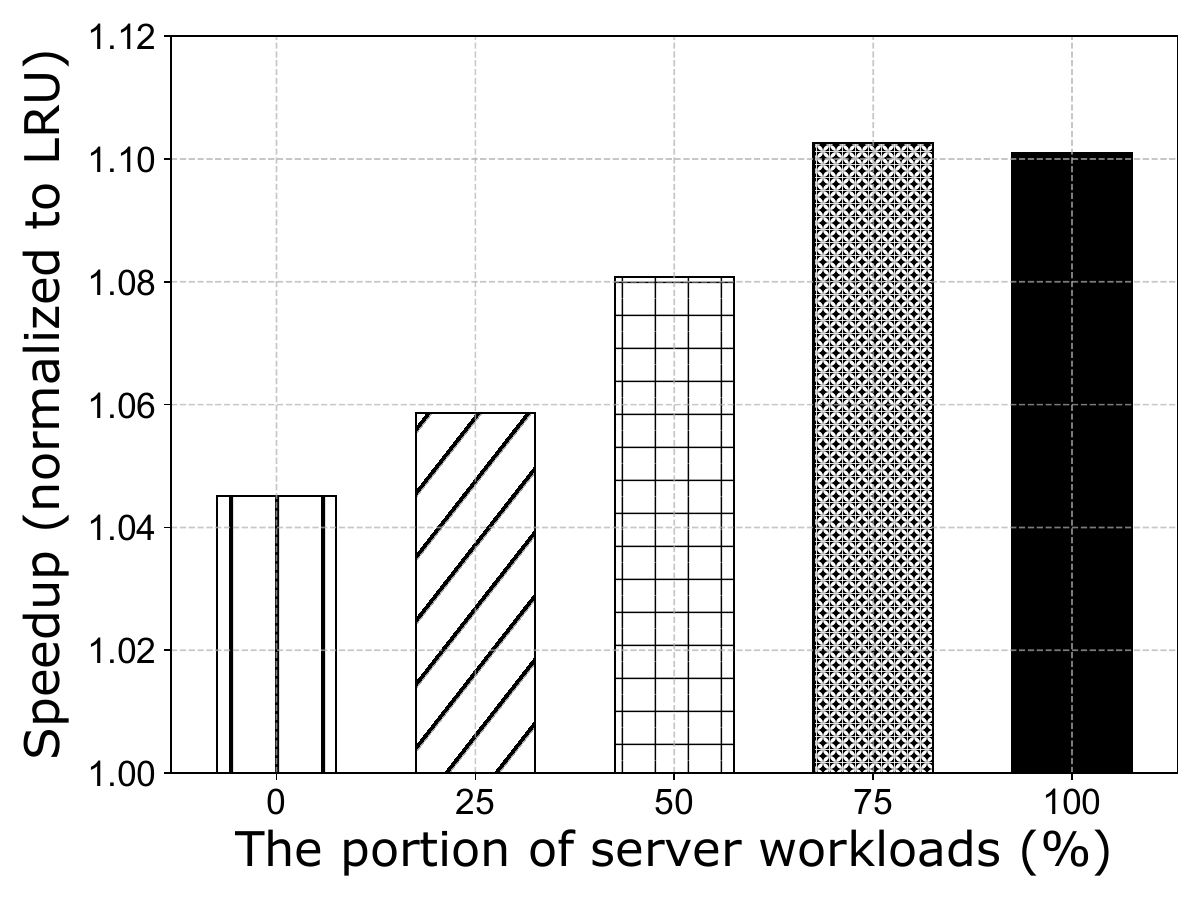}
        \vspace{-2mm}
      {\footnotesize  \caption{Mixture of benchmark suites.}}
        \label{fig:mix}
    \end{subfigure}
    \hspace{-0.3mm}
    \begin{subfigure}[t]{0.48\linewidth}
        \centering
        \includegraphics[width=0.9\linewidth]{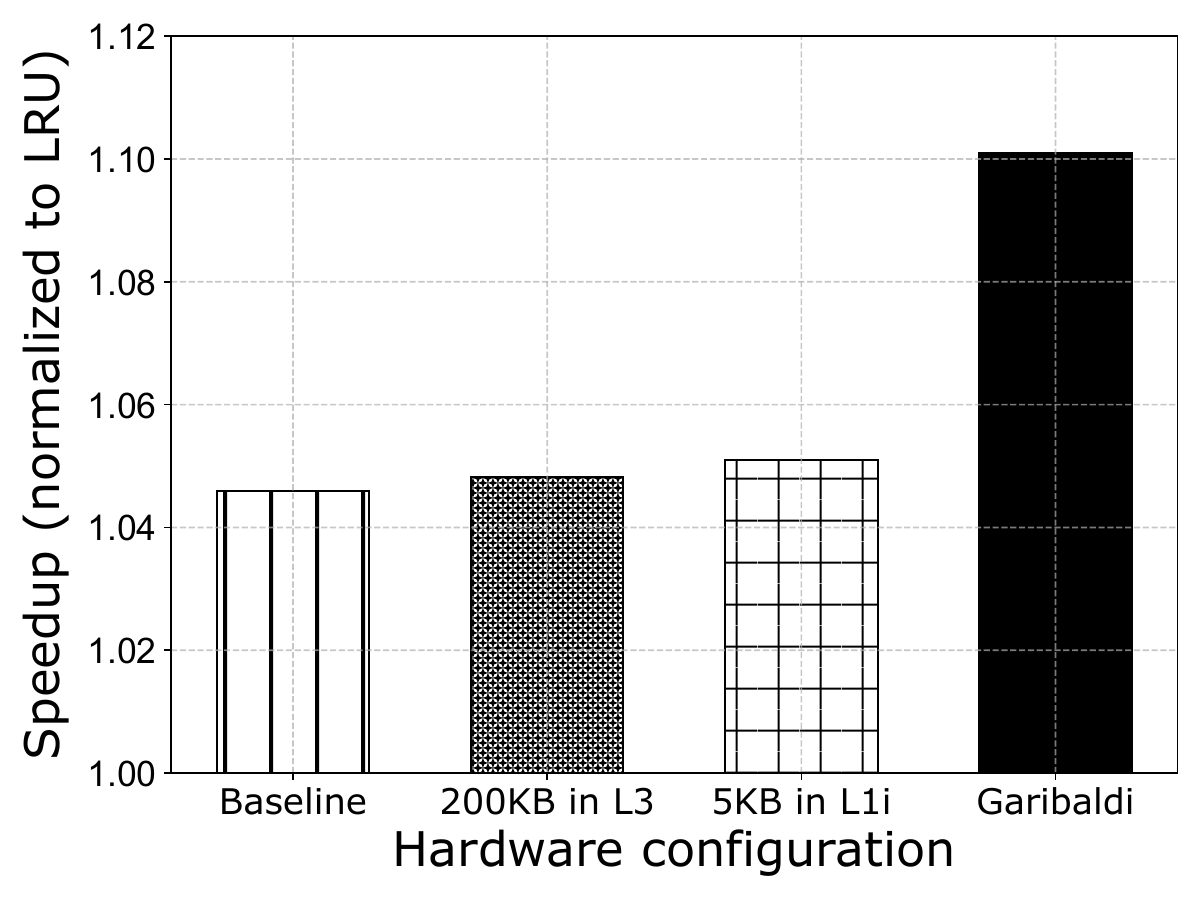}
        \vspace{-2mm}
        {\footnotesize \caption{Additional cache capacities.}}
        \label{fig:more_capacity}
    \end{subfigure}
    \vspace{-4mm}
    \caption{Evaluation results showing speedups under different configurations.}
    \label{fig:design_space_idiot}
\end{figure}



\begin{figure*}[ht]
    \includegraphics[width=0.87\linewidth]{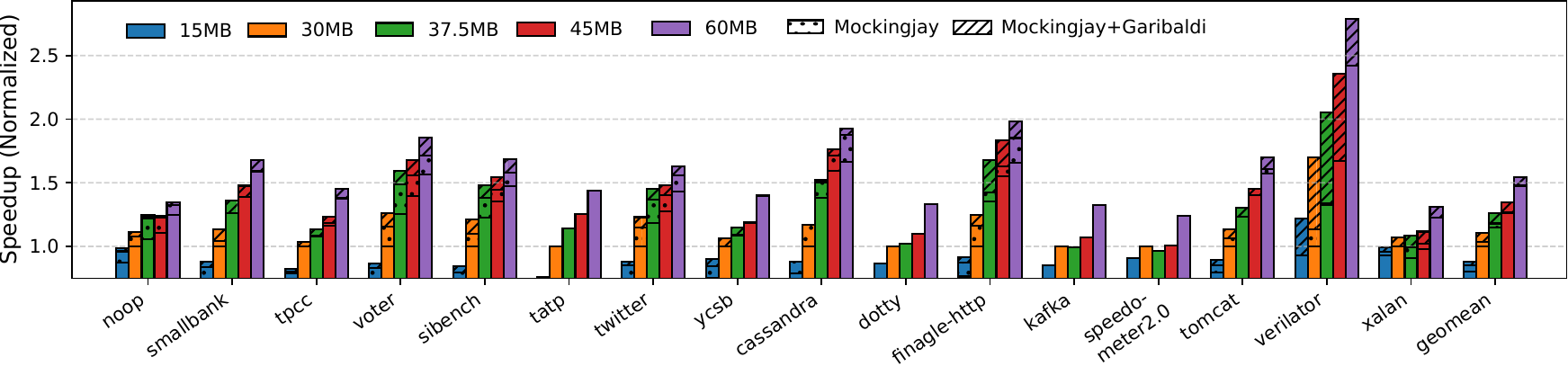}
    \vspace{-2mm}
    \caption{Performance comparison of Garibaldi+Mockingjay, Mockingjay, and LRU under varying LLC capacities, normalized to the baseline (30MB) with LRU. Associativity is fixed at 12 ways.}
    \label{fig:dse_cap}
\end{figure*} 

\begin{figure*}[ht]
    \centering
    \includegraphics[width=0.87\linewidth]{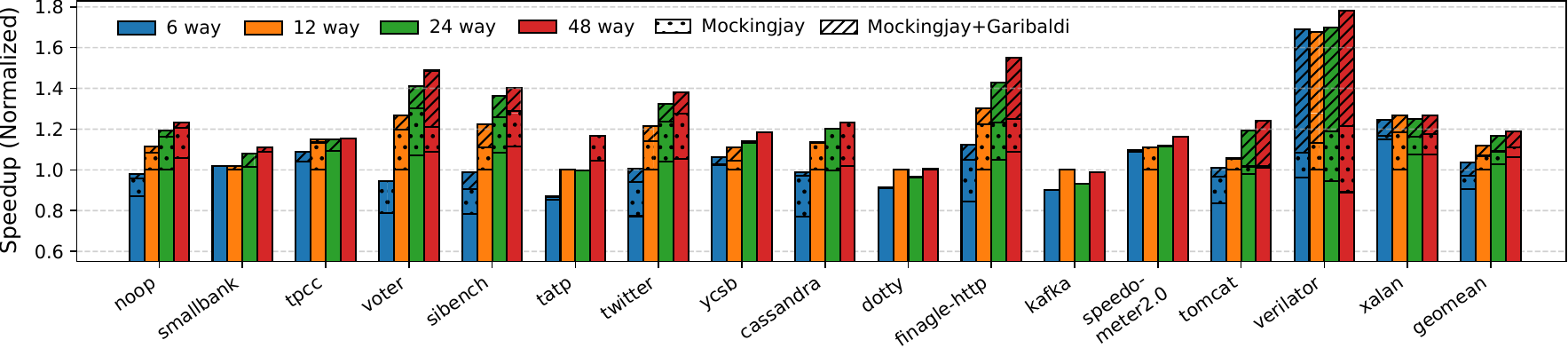}
    \vspace{-2mm}
    \caption{Performance comparison of Garibaldi+Mockingjay, Mockingjay, and LRU under varying LLC associativities, normalized to the baseline (12-way) with LRU. Capacity is fixed at 30MB.}
    \vspace{-2mm}
    \label{fig:dse_way}
\end{figure*}

\noindent \textbf{Other Comparisons:}\rv{revc}{C-7} In Fig. \ref{fig:design_space_idiot}(a) shows results for mixed server and SPEC workloads. Mockingjay over LRU consistently improves performance by 4.4--4.7\% across all configurations. With 0\% server workload, Garibaldi improves performance by 4.5\% over LRU---only a 0.11\% gain over Mockingjay (which lacks instruction victim handling). However, as more instructions become victims, Garibaldi’s benefit increases, saturating at 10.2\% (5.3\% over Mockingjay) with 75\% server workload. 
\rv{reve}{E-3} Fig. \ref{fig:design_space_idiot}(b) compares Garibaldi with configurations that allocate an extra 200KB of storage directly to the LLC and I1-i, which yield improvements of only 0.21\% and 0.48\% over Mockingjay, respectively, compared to Garibaldi’s 5.25\%.

\vspace{3mm}
\subsection{Sensitivity Studies}
%
\noindent\textbf{Sensitivity Studies of LLC Capacity:}\label{casestudy}
%
%
We evaluate the performance of Garibaldi+Mockingjay, Mockingjay, and LRU across varying LLC capacities. Fig. \ref{fig:dse_cap} shows weighted speedups normalized to the 30MB LRU baseline for 16 server workloads. Dotted stacks represent Mockingjay’s improvement, and dashed stacks indicate Garibaldi’s additional gains. Associativity is fixed at 12-way.

Mockingjay’s data caching benefits decrease with larger cache sizes. For example, in \texttt{voter}, Mockingjay achieves a 17.1\% speedup over LRU at 15MB but only 8.1\% at 60MB. On average, Mockingjay’s improvements are significant at smaller sizes (e.g., 15MB) but diminish as cache size increases, becoming negligible beyond 45MB (1.5$\times$ the baseline).
Instruction victimization sensitivity to LLC capacity varies by workload but is not eliminated entirely. For instance, in \texttt{verilator} and \texttt{finagle-http}, instruction victims decrease with larger caches, narrowing Garibaldi’s advantage. However, in workloads like \texttt{smallbank}, \texttt{tpcc}, \texttt{voter}, and \texttt{sibench}, Garibaldi consistently provides modest gains. At 60MB, Garibaldi achieves a 4.6\% improvement over LRU, while Mockingjay shows no measurable improvement. 

\noindent\textbf{Sensitivity Studies of LLC Associativity:}
We evaluate Garibaldi+ Mockingjay, Mockingjay, and LRU performance under varying LLC associativities. Fig. \ref{fig:dse_way} shows the speedup of 16 server workloads normalized to the baseline (12-way) using LRU. 

Mockingjay’s benefits generally grow with higher associativity in workloads like \ttt{noop}, \ttt{sibench}, and \ttt{twitter}. For example, in \ttt{noop}, Mockingjay achieves a 9.1\% speedup over LRU at 6-way, rising to 13.7\% at 48-way. Conversely, in \ttt{voter}, \ttt{finagle-http}, and \ttt{tomcat}, Mockingjay’s gains shrink as associativity increases due to instruction victimization, which limits its effectiveness.
Garibaldi mitigates this limitation, delivering significant additional gains as associativity increases. For instance, in \ttt{finagle-http}, at 6-way, Garibaldi+Mockingjay achieves a 32.6\% speedup over LRU, with Mockingjay contributing 23.8 percentage points (pp) and Garibaldi adding 8.8 pp. At 48-way, the combined improvement rises to 42.2\%, with Mockingjay contributing 14.5 pp and Garibaldi 27.7 pp.
These results emphasize Garibaldi’s role in addressing instruction victimization and complementing Mockingjay. On average, Garibaldi’s advantage over Mockingjay peaks at 48-way associativity (7.1\%), while Mockingjay’s improvement over LRU is smallest at 48-way (4.2\%).

\section{Conclusion}
Garibaldi introduces a pairwise instruction-data management scheme that conveys the hotness of data cachelines to corresponding instruction lines, mitigating instruction misses in shared LLCs. By protecting high-cost instructions, it addresses the “instruction victim” problem, achieving up to 13.2\% speedup over baseline and outperforming SOTA LLC policies like Mockingjay.  


\begin{acks}
  We thank the anonymous reviewers for their insightful feedback. This work was supported in part by the National Research Foundation of Korea (NRF) grant funded by the Korea government (MSIT)(RS-2024-00357037), and by Samsung Electronics Co., Ltd. (IO201210-07936-01). Won Woo Ro is the corresponding author.

\end{acks} 






\bibliographystyle{ACM-Reference-Format}
\bibliography{refs.bib}











\end{document}